\documentclass[fleqn,usenatbib,useAMS]{mnras}

\usepackage[british]{babel}
\usepackage[T1]{fontenc}
\usepackage{ae,aecompl}

\usepackage{graphicx}	
\usepackage{amsmath}	
\usepackage{amssymb}	
\usepackage{xspace}
\usepackage{verbatim}
\usepackage{color}

\newcommand{\galform}{{\sc galform}\xspace}
\newcommand{\mbh}{M_{\mathrm{BH}}}
\newcommand{\lbol}{L_{\mathrm{bol}}}
\newcommand{\rwarp}{R_{\mathrm{warp}}}
\newcommand{\ledd}{L_{\mathrm{Edd}}}

\newcommand{\angstrom}{\mbox{\normalfont\AA}}

\title[SMBH and AGN evolution]{The evolution of SMBH spin and AGN luminosities for $z<6$ within a semi-analytic model of galaxy formation.}

\author[A. J. Griffin et al.]{
Andrew J. Griffin,$^{1}$\thanks{E-mail: andrew.j.griffin@durham.ac.uk (AJG)}
Cedric G. Lacey,$^{1}$
Violeta Gonzalez-Perez,$^{1,2}$\newauthor
Claudia del P. Lagos,$^{3,4}$
Carlton M. Baugh,$^{1}$
Nikos Fanidakis$^{5,6}$
\\
$^{1}$Institute for Computational Cosmology, Department of Physics, University of Durham, South Road, Durham, DH1 3LE, UK\\
$^{2}$Institute of Cosmology and Gravitation, University of Portsmouth, Burnaby Road, Portsmouth PO1 3FX, UK\\
$^{3}$International Centre for Radio Astronomy Research, University of Western Australia, 35 Stirling Highway, Crawley, WA 6009, Australia\\
$^{4}$ARC Centre of Excellence for All Sky Astrophysics in 3 Dimensions (ASTRO 3D)\\
$^{5}$Max-Planck-Institute for Astronomy, K\"onigstugl 17, D-69117 Heidelberg, Germany\\
$^{6}$BASF, Carl-Bosch Strasse 38, 67056 Ludwigshafen, Germany
}

\date{Accepted XXX. Received YYY; in original form ZZZ}

\pubyear{2018}

\begin{document}
\label{firstpage}
\pagerange{\pageref{firstpage}--\pageref{lastpage}}
\maketitle

\begin{abstract}
Understanding how Active Galactic Nuclei (AGN) evolve through cosmic time allows us to probe the physical processes that control their evolution. 
We use an updated model for the evolution of masses and spins of supermassive black holes (SMBHs), coupled 
to the latest version of the semi-analytical model of galaxy formation \galform
using the \emph{Planck} cosmology and a high resolution
Millennium style dark matter simulation to make predictions for AGN and SMBH properties for $0 < z < 6$. 
We compare the model to the observed black hole mass function and the SMBH versus galaxy bulge mass relation at $z=0$, 
and compare the predicted bolometric, hard X-ray, soft X-ray and optical AGN luminosity functions to
observations at $z < 6$, and find that the model is in good
agreement with the observations. The model predicts that at $z<2$ and $\lbol < 10^{43} \mathrm{ergs^{-1}}$, the AGN luminosity function 
is dominated by objects accreting in an Advection Dominated Accretion Flow (ADAF) disc state, while at higher redshifts and higher luminosities 
the dominant contribution is from objects accreting via a thin disc or at super-Eddington rates. The model also predicts that 
the AGN luminosity function at $z<3$ and $\lbol < 10^{44} \mathrm{ergs^{-1}}$ is dominated by the contribution from AGN fuelled by quiescent hot halo 
accretion, while at higher luminosities and higher redshifts, the AGN luminosity function is dominated by the contribution from AGN 
fuelled by starbursts triggered by disc instabilities.
We employ this model to predict the evolution of SMBH masses, Eddington ratios, and spins, 
finding that the median SMBH spin evolves very little for $0<z<6$.
\end{abstract}




\section{Introduction}

Ever since quasars were first identified to be cosmological sources \citep{schmidt68}, a key aim has been to
to understand their evolution through cosmological time. 
Early studies showed that the number density of quasars shows strong evolution,
with more luminous quasars present at $z \approx 2$ than at $z \approx 0$, leading to the suggestion that
quasars evolve by `pure luminosity evolution' (PLE). In this scenario, quasars are long lived and 
fade through cosmic time, leading to an evolution in the luminosity function of only the characteristic luminosity \cite[e.g.][]{boyle90}.
However, more recent optical surveys, which can probe both the faint and bright end of the luminosity function, have shown not only that the slope of
the luminosity function evolves \citep[e.g.][]{richards06,croom09b}, but also that the number density
decreases at high redshift \citep[e.g.][]{fan01a,jiang16}. Surveys at X-ray wavelengths, show an evolution in the shape of
the luminosity function \citep[e.g.][]{ueda14} as well as differences between the absorbed and unabsorbed
populations \citep[e.g.][]{aird15, georgakakis15}.
Clearly, the full picture of supermassive black holes (SMBHs) and Active Galactic Nuclei (AGN) evolving through cosmological
time is complicated, and requires detailed investigation. Theoretical models and cosmological simulations have allowed us to
try to quantify the role of different contributing black hole fuelling mechanisms
(e.g. mergers, disc instabilities) and obscuration to the AGN luminosity function \citep[e.g.][]{fani12, hirschmann12},
but we do not yet fully understand the reasons for the different features of the evolution. 

The evolution of AGN through time also has significance for galaxy formation,
since AGN are thought to have a dramatic effect on their host galaxies. 
The relativistic jets from AGN can have a strong effect on the surrounding hot gas by forming huge X-ray cavities 
\citep[e.g.][]{forman05,randall11,blanton11}, or the AGN can drive powerful high-velocity outflows
\citep[e.g.][]{pounds03,reeves03,rupke11}.

The precise physical mechanism for the production of AGN jets has not yet been determined,
but the two most popular mechanisms are either that the accretion flow determines the jet power 
\citep{blandfordpayne82} or that the spin of the black hole determines the jet power \citep{blandfordznajek77}.
Many simulations of black hole accretion discs have been conducted to study jet formation, where black hole spin often plays a key role 
\citep[e.g.][]{kudoh98, hawleybalbus02,mckinney05,hawleykrolik06,tchekhovskoy12, sadowski13}. 
The importance of black hole spin has motivated  
observational studies to constrain black hole spin values
\citep{brenneman06, chiang11, done13} and cosmological simulations of black hole spin evolution
\citep{berti08, lagos09, fani11, barausse12, dotti13, fiacconi18}. The latter have been used to try to understand the role of 
black hole mergers and accretion across cosmological time on SMBH growth and spin evolution. 

Constructing a model that accounts for all the processes involved in changing SMBH spin is not a simple task - especially given
the vast range of scales involved. On sub-parsec scales, the spin of the black hole affects the radius of the last stable
orbit for orbiting material, and hence the radiative efficiency of the black hole. The SMBH spin can be misaligned with that of the accretion disc, causing 
the accretion disc to become warped, which affects how gas is accreted \citep{bardeenpetterson75}. 
On larger scales it is currently unclear
whether the gas accretes in an ordered manner or in a series of randomly oriented events \citep{kph08}, which affects how much the black hole spins up. 
On galaxy-wide scales, simulations and theoretical models predict that the SMBH may accrete gas from
cold, infalling gas made available by galaxy mergers \citep[e.g.][]{dimatteo05} or by disc instabilities \citep[e.g.][]{younger08}, or from gas accreted from the hot halo
gas surrounding the galaxy \citep[e.g.][]{bower06}. Cold streams may also supply gas to the central regions of galaxies \citep[e.g.][]{khandai12}.

To be able to model all of these processes operating over all of these scales, some form of `sub-grid' prescription is required
to be able to model the effects going on below the numerical resolution of the calculation. Therefore, investigating SMBH 
spin evolution and mass growth and exploring their effects on galaxy wide scales is well suited to using a semi-analytic
model of galaxy formation. Using a semi-analytic model of galaxy formation coupled with a large-volume, high-resolution dark matter simulation
means that we can conduct detailed simulations within a computationally reasonable time-frame, which means that we can investigate the low
redshift Universe (by comparing to observations), and can make predictions for the high-redshift Universe, with greater accuracy than previous studies using simulations.

Semi-analytic models of galaxy formation have greatly contributed to our understanding of SMBHs and AGN in galaxy formation.
\cite{bower06} used \galform 
with an AGN feedback prescription in which heating of halo gas by relativistic jets balance radiative cooling in the most massive haloes, 
to provide a match to the galaxy luminosity 
function at a range of redshifts, highlighting the potential importance of AGN feedback on galaxy formation. \cite{malbon07} extended the \galform model of \cite{baugh05} by including SMBH growth
from mergers, cold gas accreted from starbursts and from the hot halo mode introduced in \cite{bower06} to reproduce
the quasar optical luminosity function. In \cite{fani11}, \galform was updated to include an SMBH spin evolution model
in which SMBH spin evolves during accretion of gas or by merging with other SMBHs. This model was then
compared to observed AGN luminosity functions in the redshift range $0 < z < 6$ for optical and X-ray data in \cite{fani12}.
Other semi-analytic galaxy formation models have also investigated SMBH growth and evolution
\citep[e.g.][]{lagos08, marulli08, bonoli09, hirschmann12, menci13, neistein14, enoki14, shirakata18} and studies have also been
conducted using hydrodynamical simulations \citep[e.g.][]{hirschmann14, sijacki15, rosasguevara16, volonteri16, weinberger18}.

In this paper, we present predictions for the evolution of SMBH and AGN properties in the redshift range $0<z<6$, 
using an updated prescription for the evolution of SMBH spin within the \galform semi-analytic model of galaxy formation.
We include a more detailed treatment of the obscuration and compare the model predictions to more recent observational data. In a subsequent paper we will present predictions for $z>6$.

This paper is organised as follows. 
In Section \ref{sec:model} we outline the galaxy formation model and the spin evolution model and in Section \ref{sec:model_2} we outline the calculation of AGN luminosities. In Section \ref{sec:preliminaries}
we present predictions for black hole masses and spins for the model, as well as the dependence of AGN luminosities on galaxy properties.
In Section \ref{sec:comparison} we show the evolution of 
the AGN luminosity function at different wavelengths for $0 < z < 6$. In Section \ref{sec:conclusions} we give concluding remarks.


\section{The galaxy formation and SMBH evolution model}
\label{sec:model}

\subsection{The \galform model}

To make our predictions, we use the Durham semi-analytic model of galaxy formation, \galform.
Building on the principles outlined in \cite{white78}, \cite{whitefrenk91}, and \cite{cole94}, and introduced in
\cite{cole00}, in \galform galaxies form from baryons condensing within dark matter haloes, with the assembly 
of the haloes described by the dark matter halo merger trees. 
While dark matter merger trees can be calculated using a Monte-Carlo technique
that is based on the Extended Press-Schechter theory \citep{laceycole93, cole00, parkinson08}, 
they can also be extracted from 
dark matter N-body simulations \citep{kauffmann99, helly03, jiang14}, which is the method that we follow in this paper.
The baryonic physics is then modelled using a set of coupled differential equations to track the exchange of
baryons between different galaxy components.
The physical processes modelled in \galform include: 
i) the merging of dark matter haloes,
ii) shock heating and radiative cooling of gas in haloes,
iii) star formation from cold gas,
iv) photoionisation/supernova/AGN feedback, 
v) the chemical evolution of gas and stars,
vi) galaxies merging in haloes due to dynamical friction,
vii) disc instabilities
viii) the evolution of stellar populations, and
ix) the extinction and reprocessing of stellar radiation by dust. 
For a detailed description of the physical processes involved, see 
\cite{lacey16} and references therein. 

In this paper we update the model for SMBHs and AGN presented in
\cite{fani11}, superceding the equations in that paper, which
contained some typographical errors, and also putting special emphasis
on improving the model for the obscuration of AGN at X-ray and optical
wavelengths.  We incorporate the updated \cite{fani11} SMBH model in
the \cite{lacey16} \galform model as updated for the Planck-Millennium simulation by \cite{baugh18}.  The \cite{lacey16} model brings
together several \galform developments into a single model, which fits
well a wide range of observational data covering wavelengths from the
far-UV to the sub-mm in the redshift range $0<z<6$. The \cite{lacey16}
\galform model differs in a number of ways from that used in
\cite{fani11,fani12}, including having different IMFs for quiescent
and starburst star formation, as opposed to the single IMF used in
\cite{fani11}. 

The dark matter simulation used for these predictions is a new $\mathrm{(800 Mpc)^3}$ Millennium
style simulation \citep{springel05b} with cosmological parameters consistent with the \emph{Planck}
satellite results \citep{planck14} - henceforth referred to as the P-Millennium \citep{baugh18}.
The P-Millennium has an increased number of snapshots output - 270 instead of 64 for the Millennium simulation, 
the time interval between outputs has been chosen to ensure there are sufficient output snapshots for convergence of galaxy properties 
\citep[c.f.][]{benson12}. The halo mass resolution 
is $2.12 \times 10^9 h^{-1} \mathrm{M_{\odot}}$, compared to the 
halo mass resolution of $1.87 \times 10^{10} h^{-1} \mathrm{M_{\odot}}$ for the dark matter simulation
used in \cite{lacey16}. 
This halo mass resolution is a result of P-Millennium having a dark matter particle mass of $1.06 \times 10^{8} h^{-1} \mathrm{M_{\odot}}$.

\begin{table*}
\caption{The cosmological and galaxy formation parameters for this study that have been changed from the \protect\cite{lacey16} \galform model. These parameters in this study are as in \protect\cite{baugh18}. $\Omega_{\rm m0}$, $\Omega_{\rm v0}$ and $\Omega_{\rm b0}$ are the present-day density parameters in matter, vacuum energy and baryons, $h$ is the present-day Hubble parameter in units of $100 \, {\rm kms^{-1} Mpc^{-1}}$ and $\sigma_8$ is the normalization of the initial power spectrum of density fluctuations. The parameters $\gamma_{\mathrm{SN}}$ and $\alpha_{\mathrm{ret}}$ are related to supernova feedback and the return timescale for ejected gas, as described in \S3.5.2 in \protect\cite{lacey16}.}
\begin{tabular}{ |c|c|c|c| } 
\hline
Parameter & Description & \cite{lacey16} & This study \\ 
\hline
$\Omega_{\rm m0}$ & Matter density & 0.272 & 0.307 \\
\hline
$\Omega_{\rm v0}$ & Vacuum energy density & 0.728 & 0.693 \\
\hline
h & Reduced Hubble parameter & 0.704 & 0.678 \\
\hline
$\Omega_{\rm b0}$ & Baryon density & 0.0455 & 0.0483 \\
\hline 
$\sigma_{8}$ & Power spectrum normalization & 0.818 & 0.829 \\
\hline
\hline
$\alpha_{\mathrm{ret}}$ & Gas reincorporation timescale & 0.64 & 1.0 \\
\hline
$\gamma_{\mathrm{SN}}$ & Slope of SN feedback mass loading & 3.2 & 3.4 \\
\hline
& Galaxy merger timescale & \cite{jiang08} & \cite{simhacole16} \\
\hline
\end{tabular}
\label{tab:pmill_params}
\end{table*}

Because of the changed cosmological parameters and improved halo mass resolution in P-Millennium
compared to the simulation used in \cite{lacey16}, 
it was necessary to re-calibrate some of the
galaxy formation parameters - which was done in \cite{baugh18}. The new model also includes a more accurate calculation of the timescale
for galaxies to merge within a halo \citep{simhacole16}. Only two \galform parameters were changed, both relating to supernova feedback and the return of ejected gas. The 
parameters were changed in \cite{baugh18} from the \cite{lacey16} model values are shown in Table \ref{tab:pmill_params}.
This P-Millennium based model has already been used in \cite{cowley18} to make predictions for galaxies for JWST
in near- and mid-IR bands, and a model using P-Millennium and the model of \cite{gonzalezperez18} was used to study 
the effect of AGN feedback on halo occupation distribution models in \cite{mccullagh17}.

We emphasize that the aim in this paper is to
  study SMBH and AGN evolution in the framework of an existing galaxy
  formation model calibrated on a wide range observational data on
  galaxies. Therefore we do not consider any modifications to the
  underlying galaxy formation model, only to the modelling of SMBH and
  AGN within it.

\subsection{SMBH growth}
\label{sec:smbh_growth}

SMBHs in \galform grow in three different ways. 

\subsubsection{Starburst mode gas accretion}

Firstly, SMBHs can accrete gas during starbursts, which are triggered by either galaxy mergers or disc instabilities. In both of these cases, all of the remaining cold gas in a galaxy is consumed in a starburst and a fixed fraction of the mass of stars formed from the starburst feeds the SMBH, such that the accreted mass $M_{\mathrm{acc}}$ is $f_{\mathrm{BH}} M_{\star, \mathrm{burst}}$ where $M_{\star, \mathrm{burst}}$ is the mass of stars formed in the starburst and $f_{\mathrm{BH}}$ is a free parameter \citep[c.f.][]{lacey16}. Note that the mass of the stars formed is less than the initial mass of the gas in the starburst due to the ejection of gas by supernova feedback.

A galaxy merger can cause gas to be transferred to the centre of the galaxy and trigger a burst of star formation \citep[e.g.][]{mihos96}. Some of this gas is then available to feed the central SMBH \citep{kauffmann00, malbon07}.  In the model, if the mass ratio of the two galaxies is less than 0.3, the merger is defined as a minor merger, and a starburst is triggered for a minor merger with mass ratio above 0.05. If the mass ratio of the galaxies is greater than 0.3, the merger is defined as a major merger, and a spheroid is formed. This is described in Section 3.6.1 of \cite{lacey16}.

Disc instabilities cause a bar to be formed, which disrupts the galaxy disc \citep{efstathiou82} and transfers gas to the centre of the galaxy to be fed into the SMBH. Disc instabilities driving gas into the centres of galaxies is an effect seen in various hydrodynamical simulations \citep[e.g.][]{hohl71,bournaud05,younger08}, and used as a channel of black hole/bulge growth in many semi-analytic models of galaxy formation \citep[e.g.][]{delucia11,hirschmann12,menci14,croton16,lagos18}, although the implementation of these disc instabilities varies between models.  Most models use the disc instability criterion of \cite{efstathiou82} in which the disc becomes unstable if it is sufficiently self-gravitating, however different models use this condition differently.  For example, in the model of \cite{hirschmann12}, if a disc is unstable, then enough gas and stars are transferred from the disc to the bulge to completely stabilise the disc, while in \galform, we assume that if a disc is unstable, then it is completely destroyed and forms a bulge. Numerical simulations of isolated disks show that disc instabilities can transfer large fractions of gas and stars into the bulge in some situations \citep[e.g.][]{bournaud07,elmegreen08,saha18}.
Disc instabilities in \galform are described in Section 3.6.2 of
\cite{lacey16}.

\subsubsection{Hot halo mode gas accretion}

In \galform, we assume that SMBHs can also accrete gas from the hot
gas atmospheres of massive haloes: when large haloes collapse, gas is
shock heated to form a quasistatic hot halo atmosphere. For
sufficiently massive haloes, the cooling time of this gas is longer
than its free-fall time, and the SMBH is fed with a slow inflow from
the halo's hot atmosphere - `hot halo mode accretion'
\citep{bower06}. The black hole is assumed to grow by this fuelling
mechanism only when AGN feedback is operational. In this regime,
energy input by a relativistic jet is assumed to balance radiative
cooling in the halo, with the mass accretion rate onto the black hole
$\dot{M}$ being determined by this energy balance condition. The mass
$M_{\mathrm{acc}}$ accreted onto the SMBH in a simulation timestep $\Delta
t_{\mathrm{step}}$ is then $\dot{M} \Delta t_{\mathrm{step}}$.
The hot halo accretion mode is fully described in Section 3.5.3 of
\cite{lacey16}.

\subsubsection{SMBH mergers}

SMBHs can be built up by SMBH-SMBH mergers. When galaxies merge, dynamical friction from gas, stars and dark matter
causes the SMBH of the smaller galaxy to sink towards the other SMBH. Then, as the separation decreases, gravitational radiation
provides a mechanism by which the SMBHs can lose angular momentum and spiral in to merge and form a larger SMBH. 
In the model, we assume the timescale on which the SMBHs merge is short, so that the SMBHs merge when the galaxies merge.

\subsection{SMBH seeds}

The starting point for the treatment of SMBHs in the model is SMBH seeds that eventually grow by accretion of gas and 
by merging with other SMBHs to form the objects in the Universe today. The processes for SMBH seed formation are 
uncertain (see e.g. \citealt{volonteri10}, and references therein) and so we simply add a seed SMBH of mass
$M_{\mathrm{seed}}$ into each halo, where $M_{\mathrm{seed}}$ is a parameter that we can vary. Unless otherwise stated,
this parameter has the value $M_{\mathrm{seed}} = 10 h^{-1} M_{\odot}$ - representative of the SMBH seeds formed by 
stellar collapse. The effect of varying this seed mass is discussed in Appendix \ref{app:seed_mass}.

\subsection{SMBH mass growth and spinup by gas accretion}
\label{sec:gas_accretion}

Our model includes the evolution of SMBH spin. In this model, SMBHs can change spin in two ways:
(i) by accretion of gas or (ii) by merging with another SMBH.
The SMBH spin is characterised by the dimensionless spin parameter, $a = cJ_{\mathrm{BH}}/GM_{\mathrm{BH}}^{2}$, 
within the range $-1 \leq a \leq 1$,
where $J_{\mathrm{BH}}$ is the angular momentum of the SMBH, and $M_{\mathrm{BH}}$ is the mass of the SMBH. 
$a=0$ represents a black hole that is not spinning and $a=1$ or $a=-1$ represents a maximally
spinning black hole. The sign of $a$ is defined by the direction of the angular momentum of the black hole 
relative to that of the innermost part of the accretion disc, so for $a>0$ the black hole is spinning in the same direction
as the inner accretion disc and for $a<0$ the black hole is spinning in the opposite direction to the inner accretion disc.
To calculate the SMBH spin, $a^{f}$ after an accretion episode, we use the expression in 
\cite{bardeen70}\footnote{Note that equation (\ref{eq:a_f}) is corrected from \cite{fani11} equation (6).}:

\begin{equation}
 a^{f} = \frac{1}{3} \sqrt{\hat{r}_{\mathrm{lso,i}}} \frac{M_{\mathrm{BH,i}}}{M_{\mathrm{BH,f}}} \Big( 4 - \Big[ 3\hat{r}_{\mathrm{lso,i}} \Big( \frac{M_{\mathrm{BH,i}}}{M_{\mathrm{BH,f}}} \Big)^{2} - 2\Big]^{1/2} \Big), \label{eq:a_f}
\end{equation}

\noindent
where $\hat{r}_{\mathrm{lso}}$ is the radius of the last stable 
circular orbit in units of the gravitational radius, $R_{\rm{G}} = G M_{\mathrm{BH}} / c^2$, 
and the subscripts i and f indicate values at the start and end of an accretion event. The black hole mass before and after an accretion event are related by:

\begin{equation}
 M_{\mathrm{BH,f}} = M_{\mathrm{BH,i}} + (1-\epsilon_{\mathrm{TD}})\Delta M,
\end{equation}

\noindent
where $\Delta M$ is the mass accreted from the disc in this accretion episode and $\epsilon_{\mathrm{TD}}$, 
the radiative accretion efficiency for a thin accretion disc, is given by: 

\begin{equation}
 \epsilon_{\mathrm{TD}} = 1 - \Big( {1- \frac{2}{3 \hat{r}_{\mathrm{lso}}}} \Big)^{1/2}. \label{eq:e_td}
\end{equation}

\noindent
$\hat{r}_{\mathrm{lso}}$ is calculated from the spin $a$, as in \cite{bardeen72}:

\begin{equation}
 \hat{r}_{\mathrm{lso}} = 3 + Z_{2} \mp \sqrt{(3 - Z_{1})(3 + Z_{1} + 2 Z_{2})},
\end{equation}

\noindent
with the minus sign for $a>0$ and the positive sign for $a<0$. The functions $Z_{1}$ and $Z_{2}$
are given by:

\begin{equation}
 Z_{1} = 1 + (1 - |a|^2)^{1/3}[(1 + |a|)^{1/3} + (1 - |a|)^{1/3}],
\end{equation}

\begin{equation}
 Z_{2} = \sqrt{3|a|^2 + Z_{1}^2}.
\end{equation}

\begin{figure}
\centering
\includegraphics[width=.8\linewidth]{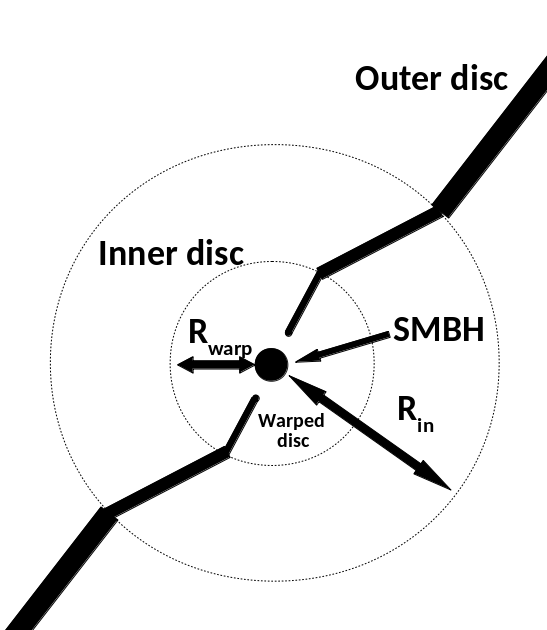}
\caption{A diagram showing the various scales involved in the gas accretion - the warp radius $\rwarp$ and
the inner radius $R_{\mathrm{in}}$. We refer to the region within $R_{\mathrm{in}}$
as the inner disc and the region outside of $R_{\mathrm{in}}$ as the outer disc.}
\label{fig:accretion_disc_scenario} 
\end{figure}

We consider the accretion disc in three separate parts as shown in Figure \ref{fig:accretion_disc_scenario} - an outer disc at radii greater
than an inner radius, $R_{\mathrm{in}}$, an inner disc for radii less than $R_{\mathrm{in}}$, and a warped disc for radii
less than the warp radius, $R_{\mathrm{warp}}$. 
The SMBH has an angular momentum $\vec{J}_{\mathrm{BH}}$, and the angular momentum of the disc within $R_{\mathrm{in}}$ is $\vec{J}_{\mathrm{in}}$. If $\vec{J}_{\mathrm{BH}}$ is not in the same direction as $\vec{J}_{\mathrm{in}}$ a spinning black hole induces a Lense-Thirring precession in the misaligned disc elements.
Because the precession rate falls off as $R^{-3}$, at smaller radii the black hole angular momentum
and the accretion disc angular momentum vectors will become exactly aligned or anti-aligned, whereas at sufficiently
large radii there will still be a misalignment \citep{bardeenpetterson75}.
The transition between these two regions occurs at the so-called `warp radius', $R_{\mathrm{warp}}$.
The angular momentum of the disc within the warp radius is $\vec{J}_{\mathrm{warp}}$.
At the start of an accretion event, the angular momentum $\vec{J}_{\mathrm{warp}}$ within $R_{\mathrm{warp}}$ is assumed to be aligned with $\vec{J}_{\mathrm{in}}$.
As a result of the torques, $\vec{J}_{\mathrm{BH}}$ then aligns with 
$\vec{J}_{\mathrm{tot}} = \vec{J}_{\mathrm{BH}} + \vec{J}_{\mathrm{warp}}$ 
(which remains constant during this alignment process) and 
$\vec{J}_{\mathrm{warp}}$ either anti-aligns or aligns with $\vec{J}_{\mathrm{BH}}$ \citep{kinglub05}. The gas within $R_{\mathrm{warp}}$ is then assumed to be accreted onto the SMBH from the aligned/anti-aligned disc.
As more gas is accreted, $\vec{J}_{\mathrm{BH}}$ eventually aligns with the rest of the inner disc, as the 
gas in the inner disc is consumed.

We consider two alternative scenarios for how the angular momentum directions of the inner and outer disc are related. 
In the `prolonged mode' accretion scenario, the angular momentum of the inner disc is in the same direction as the angular 
momentum of the outer disc, $\vec{J}_{\mathrm{out}}$, but in the the `chaotic mode' accretion scenario 
introduced in \cite{kph08}, the orientation of the angular momentum of the inner disc is randomly oriented  
with respect to the angular momentum of the outer disc. \cite{kph08} propose that $R_{\mathrm{in}}$ is the
self-gravity radius of the disc, and we assume this in our model. 

The motivation for chaotic mode accretion is
twofold. Firstly, the \cite{soltan82} argument, a comparison of the integral of the quasar luminosity function
over luminosity and redshift to the integral over the black hole mass function in the local Universe, implies an average radiative efficiency of
SMBH growth of $\epsilon \approx 0.1$ (which corresponds to a spin value of $a \approx 0.67$), suggesting that SMBHs in the Universe are typically not maximally spinning, as 
we would expect from SMBHs that have been spun up by the accretion of gas that is aligned in the same direction, 
as in the prolonged accretion scenario. Secondly,
AGN jets seem to be misaligned with their host galaxies \citep[e.g.][]{kinney00,sajina07}, 
suggesting a misaligned accretion of material onto the SMBH.

Accretion continues in this manner until the gas in the outer disc has been consumed. For this analysis,
we adopt chaotic mode accretion as our standard choice.

\subsection{Warped accretion discs}

To obtain the warp radius, $R_{\mathrm{warp}}$, of an accretion disc, we need expressions for the structure of the accretion disc.
There are two different types of accretion discs: i) physically thin, optically thick, radiatively efficient `thin discs' 
\citep{shakurasunyaev73} and ii) physically thick, optically thin, radiatively inefficient Advection Dominated Accretion Flows
\citep[ADAFs - see][for a review]{yn14}. 
\cite{shakurasunyaev73} introduced the `$\alpha$-prescription' to solve the accretion disc equations for a 
thin disc, 
where the viscosity, $\nu$, is given by
$\nu=\alpha_{\mathrm{TD}} c_{s} H$, where $\alpha_{\mathrm{TD}}$ is the dimensionless \cite{shakurasunyaev73} parameter, $c_{s}$ is the sound speed and $H$ is the disc semi-thickness.
In this analysis, we use the solutions of \cite{collin90}, in which the accretion disc equations are solved for AGN discs,
assuming this $\alpha$-prescription.
We use their solution for the regime where the opacity is dominated by electron scattering and where
gas pressure dominates over radiation pressure.

The disc surface density, $\Sigma$, is then given by:

\begin{equation}
 \Sigma = 6.84 \times 10^{5} \, \mathrm{g} \, \mathrm{cm^{-2}} \, \alpha_{\mathrm{TD}}^{-4/5} \dot{m}^{3/5} \Big( \frac{\mbh}{10^{8} M_{\odot}} \Big)^{1/8}
\Big( \frac{R}{R_{\mathrm{S}}} \Big)^{-3/5},
\end{equation}

\noindent
where $\dot{m} = \dot{M} / \dot{M}_{\mathrm{Edd}}$ is the dimensionless mass accretion rate, 
$R$ is the radius from the centre of the disc and $R_{\mathrm{S}} = 2G \mbh / c^2$ is the Schwarzschild radius. 
The value we use for $\alpha_{\mathrm{TD}}$ is given in Table \ref{tab:free_params}.
The 
disc semi-thickness $H$ is given by\footnote{Note that equation (\ref{eq:h_r}) is different to \cite{fani11} equation (25).}:

\begin{equation} 
 \frac{H}{R} = 1.25 \times 10^{-3}  \, \alpha_{\mathrm{TD}}^{-1/10} \dot{m}^{1/5} \, \Big( \frac{\mbh}{10^{8} M_{\odot}} \Big)^{-1/10}
\, \Big( \frac{R}{R_{\mathrm{S}}} \Big)^{1/20}. \label{eq:h_r}
\end{equation}

We calculate the Eddington luminosity using:

\begin{equation}
 \ledd = \frac{4 \pi G \mbh c}{\kappa} = 1.26 \times 10^{46} \, \Big( \frac{\mbh}{10^8 M_{\odot}} \Big) \, \rm{ergs^{-1}},
\end{equation}

\noindent
where $\kappa$ is the opacity, for which we have used the electron
scattering opacity for pure hydrogen gas. We calculate the Eddington
mass accretion rate $\dot{M}_{\mathrm{Edd}}$ from $\ledd$ using a
nominal accretion efficiency $\epsilon = 0.1$ \cite[as used
in][]{yn14} chosen so that the Eddington normalised mass
accretion rate $\dot{m}$ does not depend on the black hole spin:

\begin{equation}
 \dot{M}_{\mathrm{Edd}} = \frac{\ledd}{0.1 c^2}.
\end{equation}

Note that for the calculation of the luminosities, we do use the spin-dependent radiative efficiency. We then follow the method of \cite{natarajan98} and \cite{volonteri07} and take the warp radius
as the radius at which the timescale for 
radial diffusion of the warp due to viscosity is equal to the local Lense-Thirring precession timescale.
This then gives an expression for the warp radius\footnote{Note that equation (\ref{eq:r_warp}) is different to \cite{fani11} equation (15).}:

\begin{equation}
 \frac{\rwarp}{R_{S}} = 3410 \, a^{5/8} \alpha_{\mathrm{TD}}^{-1/2} \dot{m}^{-1/4} \, \Big( \frac{\mbh}{10^{8} M_{\odot}} \Big)^{1/8} 
\, \Big( \frac{\nu_{2}}{\nu_{1}} \Big)^{-5/8}, \label{eq:r_warp}
\end{equation}

\noindent
where $\nu_{1,2}$ are the horizontal and vertical viscosities respectively. For this analysis, we assume that 
$\nu_{1}=\nu_{2}$ \citep[e.g.][]{kph08}. The warp mass can then be calculated using:

\begin{equation}
 M_{\mathrm{warp}} = \int^{\rwarp}_{0} 2 \pi\Sigma(R) R^{2} dR, \label{eq:m_warp_int}
\end{equation}

\noindent
to give an expression\footnote{Note that equation (\ref{eq:m_warp}) is different to \cite{fani11} equation (18).}:

\begin{equation}
 M_{\mathrm{warp}} = 1.35 \, M_{\odot} \alpha_{\mathrm{TD}}^{-4/5}  
\dot{m}^{3/5} \, \Big( \frac{\mbh}{10^8 M_{\odot}}\Big)^{11/5} \, \Big( \frac{\rwarp}{R_{\mathrm{S}}} \Big)^{7/5}. \label{eq:m_warp}
\end{equation}

\begin{table*}
\caption{The values for the SMBH/AGN free parameters in the model. The upper part of the table shows parameters where the values 
adopted are from other studies, whereas the lower part of the table gives parameters which have been calibrated 
on the luminosity functions in Section \ref{sec:L_AGN_galaxy}.}
\begin{tabular}{|c|c|c|c|}
\hline
Parameter & \cite{fani12} & Adopted here & Significance \\
\hline
$\alpha_{\mathrm{ADAF}}$ & 0.087 & 0.1 & \cite{shakurasunyaev73} viscosity parameter for ADAFs \\
\hline
$\alpha_{\mathrm{TD}}$ & 0.087 & 0.1 & \cite{shakurasunyaev73} viscosity parameter for TDs \\
\hline
$\delta_{\mathrm{ADAF}}$ & $2000^{-1}$ & 0.2 & Fraction of viscous energy transferred to electrons in ADAF \\
\hline
$\dot{m}_{\mathrm{crit, ADAF}}$ & 0.01 & 0.01 & Boundary between thin disc and ADAF accretion \\
\hline
\hline
$\eta_{\mathrm{Edd}}$ & 4 & 4 & Super-Eddington suppression factor \\
\hline
$f_{\mathrm{q}}$ & 10 & 10 & Ratio of lifetime of AGN episode to bulge dynamical timescale \\
\hline
\end{tabular}
\label{tab:free_params}
\end{table*}

\subsection{Self-gravitating discs}

In the chaotic mode accretion scenario of \cite{kph08}, the inner radius, $R_{\mathrm{in}}$, is assumed to be equal to the 
disc self-gravity radius, $R_{\mathrm{sg}}$. The self-gravity radius of the accretion disc is the radius at which the 
vertical gravity due to the disc equals the vertical gravity of the central SMBH at the disc midplane. For thin discs and ADAFs, for thin discs (where $\dot{m}>\dot{m}_{\mathrm{crit,ADAF}}$), 
the self-gravity condition is \citep{pringle81}:

\begin{equation}
 M_{\mathrm{sg}} = \mbh \frac{H}{R}, \label{eq:m_sg_h_r}
\end{equation}

\noindent
where $M_{\mathrm{sg}}$ is the disc mass within the radius $R_{\mathrm{sg}}$. For ADAFs 
(where $\dot{m}<\dot{m}_{\mathrm{crit,ADAF}}$), $H \sim R$, so the self-gravity condition is:

\begin{equation}
 M_{\mathrm{sg}} = \mbh.
\end{equation}

Using the accretion disc solutions of \cite{collin90}, we derive an expression for the self-gravity 
radius for thin discs\footnote{Note that equation (\ref{eq:r_sg}) is different to \cite{fani11} equation (24).}:

\begin{equation}
 \frac{R_{\mathrm{sg}}}{R_{\mathrm{S}}} = 4790 \, \alpha_{\mathrm{TD}}^{14/27} \dot{m}^{-8/27} \, \Big( \frac{\mbh}{10^8 M_{\odot}}\Big)^{-26/27}, \label{eq:r_sg}
\end{equation}

\noindent
and using an integral similar to equation (\ref{eq:m_warp_int}), the self-gravity mass for the thin disc 
is given by\footnote{Note that equation (\ref{eq:m_sg}) is different to \cite{fani11} equation (26).}:

\begin{equation}
 M_{\mathrm{sg}} = 1.35 \, M_{\odot} \alpha_{\mathrm{TD}}^{-4/5}  
\dot{m}^{3/5} \, \Big( \frac{\mbh}{10^8 M_{\odot}}\Big)^{11/5} \, \Big( \frac{R_{\mathrm{sg}}}{R_{\mathrm{S}}} \Big)^{7/5}. \label{eq:m_sg}
\end{equation}

\subsection{Numerical procedure for modelling SMBH accretion}

We have calculated results for both the prolonged and chaotic scenario, and for gas accreted in increments of the self-gravity mass or warp mass. We present predictions mostly for our standard case in which mass is accreted in increments of the self-gravity mass
and assuming the chaotic mode of accretion. We find that the predicted spin distribution of the SMBHs is the same if we use increments of the self-gravity mass
or the warp mass (c.f. Figure \ref{fig:spin_vs_mbh}) and so we use increments of the self-gravity mass as it is computationally faster. This is because when gas is accreted onto the SMBH in increments of the warp mass, for small SMBHs the warp mass is very small, and so in each accretion event the SMBH grows by a very small amount in each accretion event. First, we present the numerical procedure when mass is accreted in increments of the warp mass \citep[c.f.][]{volonteri07,fani11}, and then the case where mass is accreted in increments of the self-gravity mass \citep[c.f.][]{kph08}.

\subsubsection{Accretion in increments of the warp mass}

For the first warp mass of gas, the angular momentum of the SMBH, $\vec{J}_{\mathrm{BH}}$, and 
the angular momentum of the inner disc, $\vec{J}_{\mathrm{in}}$, are assigned a random angle, $\theta_{i}$, in the 
range [0, $\pi$] radians. In the chaotic mode, each time the inner disc is consumed, $\theta_{i}$ is assigned a new random angle.
The gas with $R < R_{\mathrm{warp}}$ initially has angular momentum $\vec{J}_{\mathrm{warp}}$ 
aligned with $\vec{J}_{\mathrm{in}}$, so $\theta_{i}$ is also the initial angle between $\vec{J}_{\mathrm{BH}}$
and $\vec{J}_{\mathrm{warp}}$.
$\vec{J}_{\mathrm{BH}}$ and $\vec{J}_{\mathrm{warp}}$ are then evolved according to the Lense-Thirring effect 
described in Section \ref{sec:gas_accretion}, with $\vec{J}_{\mathrm{BH}}$ and $\vec{J}_{\mathrm{warp}}$ respectively aligning and 
aligning/anti-aligning with $\vec{J}_{\mathrm{tot}}$. 
The magnitude of $\vec{J}_{\mathrm{BH}}$ remains constant during this process, but the magnitude of $\vec{J}_{\mathrm{warp}}$ changes.
This is treated as happening before the mass consumption onto the SMBH starts.

We calculate the angular momentum of the material within the warped disc as 
$J_{\mathrm{warp}} = M_{\mathrm{warp}} \sqrt{G \mbh \rwarp}$ and the angular momentum of the black hole, 
$J_{\mathrm{BH}} = 2^{-1/2} \mbh a \sqrt{G \mbh R_{\mathrm{S}}}$.
Then the ratio of these two quantities is:

\begin{equation}
  \frac{J_{\mathrm{warp}}}{2 J_{\mathrm{BH}}} = \frac{M_{\mathrm{warp}}} {\sqrt{2} a \mbh} \Big( \frac{\rwarp}{R_{\mathrm{S}}} \Big)^{1/2}.
\end{equation}

\noindent
Whether $\vec{J}_{\mathrm{warp}}$ and $\vec{J}_{\mathrm{BH}}$ align or anti-align with each other depends on 
this ratio and on the angle $\theta_{i}$. 
If $\cos\theta_{i} > - J_{\mathrm{warp}} / 2 J_{\mathrm{BH}}$, $\vec{J}_{\mathrm{warp}}$ and $\vec{J}_{\mathrm{BH}}$ 
become aligned
(prograde accretion), whereas if $\cos\theta_{i} < - J_{\mathrm{warp}} / 2 J_{\mathrm{BH}}$, $\vec{J}_{\mathrm{warp}}$ and $\vec{J}_{\mathrm{BH}}$ 
become anti-aligned (retrograde accretion). 
The angle between $\vec{J}_{\mathrm{BH}}$ and $\vec{J}_{\mathrm{in}}$ 
after the accretion event, $\theta_{f}$, is determined by conservation of $\vec{J}_{\mathrm{tot}}$ 
and |$\vec{J}_{\mathrm{BH}}$| and is given by:

\begin{equation}
 \mathrm{cos}\theta_{f} = \frac{J_{\mathrm{warp}} + J_{\mathrm{BH}} \mathrm{cos}\theta_{i} }{\sqrt{J_{\mathrm{BH}}^{2} + J_{\mathrm{warp}}^{2} + 2 J_{\mathrm{warp}} J_{\mathrm{BH}} \mathrm{cos}\theta_{i}}}. 
\end{equation}

When a new warp mass $M_{\mathrm{warp}}$, is then consumed, the gas is given a new $\vec{J}_{\mathrm{warp}}$ pointing in the same direction
as the inner disc and the same process happens again.
This repeated process has the effect that $\vec{J}_{\mathrm{BH}}$ gradually aligns with the 
angular momentum of the inner accretion disc, $\vec{J}_{\mathrm{in}}$ as more gas is accreted. Eventually the gas in the inner disc is completely consumed.

In the prolonged mode, this process continues until all of the
gas in the outer disc has also been consumed, whereas in the
chaotic mode, once a self-gravity mass of gas has been consumed, the
angle between $\vec{J}_{\mathrm{in}}$ and $\vec{J}_{\mathrm{out}}$ is
randomised again.

\subsubsection{Accretion in increments of the self-gravity mass}

In the scenario where gas is being accreted in increments of the
self-gravity mass of gas, the above procedure is followed, but only once for each inner disc of gas consumed. For this case, the ratio of angular momenta is given by:

\begin{equation}
   \frac{J_{\mathrm{in}}}{2 J_{\mathrm{BH}}} = \frac{M_{\mathrm{sg}}} {\sqrt{2} a \mbh} \Big( \frac{\mathrm{min}(\rwarp, R_{\mathrm{sg}})}{R_{\mathrm{S}}} \Big)^{1/2}.
\end{equation}

In the future we plan a more thorough analysis of the effect on the spin evolution of accreting in increments of self-gravity mass
compared to increments of warp mass. The AGN luminosities are not affected by this choice as they depend on the accreted
mass and the SMBH spin as we describe in Section \ref{sec:lbol}.

\subsection{Spinup by SMBH mergers}

The other way in which an SMBH can change its spin is by merging with
another SMBH. The spin of the resulting SMBH depends on the spins
of the two SMBHs that merge and on the angular
momentum of their binary orbit.  To determine the final spin,
$\textbf{a}_{f}$, we use the expressions obtained from numerical
simulations of BH-BH mergers in \cite{rezz08}:

\begin{equation}
\begin{split}
 |\textbf{a}_{f}| = \frac{1}{(1+q)^{2}} \Big( |\textbf{a}_{1}^{2}| + |\textbf{a}_{2}^{2}|q^{4} + 2|\textbf{a}_{1}||\textbf{a}_{2}|q^{2} \cos\phi + \\
    2(|\textbf{a}_{1}|\cos\theta + |\textbf{a}_{2}|q^{2}\cos\xi)|\textbf{l}|q +|\textbf{l}|^{2}q^{2} \Big)^{1/2},
\end{split}
\end{equation}

\noindent
where $\textbf{a}_{1,2}$ are the spins of the SMBHs, $q$ is the mass
ratio $M_{1}/M_{2}$, with $M_{1}$ and $M_{2}$ chosen such that $q \leq
1$, $\mu$ is the symmetric mass ratio $q/(q + 1)^{2}$, and
$\textbf{l}$ is the contribution of the orbital angular momentum
to the spin angular momentum of the final black hole. It is
assumed that the direction of $\textbf{l}$ is that of the initial
orbital angular momentum, while its magnitude is given by:

\begin{equation}
\begin{split}
 |\textbf{l}| = \frac{s_{4}}{(1+q^2)^{2}} (|\textbf{a}_{1}|^{2} + |\textbf{a}|^{2}_{1}q^{4} + 2|\textbf{a}_{1}||\textbf{a}_{2}|q^{2}\cos\phi) + \\ \Big( \frac{s_{5}\mu + t_{0} + 2}{1+q^{2}} \Big) 
(|\textbf{a}_{1}|\cos\theta + |\textbf{a}_{2}|q^{2}\cos\xi) + \\ 2\sqrt{3} + t_{2}\mu + t_{3}\mu^{2}.
\end{split}
\end{equation}

\noindent
where   $s_{4}=-0.129$, $s_{5}=-0.384$, $t_{0}=-2.686$,
$t_{2}=-3.454$, $t_{3}=2.353$ are values obtained in \cite{rezz08}.
The angles $\phi$, $\theta$ and $\xi$ are the angles between the spins
of the two black holes and their orbital angular momentum, and are
given by:

\begin{equation}
 \cos\phi = \hat{\bf{a_{1}}} \cdot \hat{\bf{a_{2}}},
\end{equation}

\begin{equation}
 \cos\theta =  \hat{\bf{a_{1}}} \cdot \hat{\bf{l}},
\end{equation}

\begin{equation}
 \cos\xi = \hat{\bf{a_{2}}} \cdot \hat{\bf{l}}.
\end{equation}

\noindent
When we consider two SMBHs merging, we calculate the angles between the three different vectors
by randomly selecting directions for $\bf{a_{1}}$, $\bf{a_{2}}$ and $\bf{l}$ uniformly over the surface of a sphere. This prescription makes the assumption that the radiation of gravitational
waves does not affect the direction of the orbital angular momentum as the binary orbit shrinks, and we also assume that the mass
lost to gravitational radiation is negligible.


\section{Calculating AGN luminosities}
\label{sec:model_2}

\subsection{AGN bolometric luminosities}
\label{sec:lbol}

From the mass of gas that is accreted onto the SMBH, we can calculate a radiative bolometric luminosity as follows. 
In the starburst mode, we assume that during an accretion episode the accretion rate is constant over 
a time $f_{\mathrm{q}}t_{\mathrm{bulge}}$, where $t_{\mathrm{bulge}}$ is the dynamical timescale of the bulge
and $f_{\mathrm{q}}$ is a free parameter, given in Table \ref{tab:free_params}. Therefore, the mass accretion rate is given by:

\begin{equation}
 \dot{M} = \frac{M_{\mathrm{acc}}}{f_{q} t_{\mathrm{bulge}}}, \label{eq:mdot_macc}
\end{equation}

\noindent
where $M_{\mathrm{acc}}$ is as defined in Section \ref{sec:smbh_growth}. 
In the hot halo mode, which is only active when AGN feedback is active, the mass accretion rate is determined by the condition that the energy released is just enough to balance radiative cooling:

\begin{equation}
 \dot{M} = \frac{L_{\mathrm{cool}}}{\epsilon_{\mathrm{heat}} c^2}
\end{equation}

\noindent 
where $L_{\mathrm{cool}}$ is the radiative cooling luminosity of the hot halo gas, and $\epsilon_{\mathrm{heat}}$ is the efficiency of halo heating, which is treated as a free parameter \citep[c.f.][]{lacey16}.

We then calculate the bolometric luminosity for a thin accretion disc using:

\begin{equation}
 L_{\mathrm{bol, TD}} = \epsilon_{\mathrm{TD}} \dot{M} c^2,
\end{equation}

\noindent
where the radiative efficiency $\epsilon_{\mathrm{TD}}$ for the thin disc case depends on the black hole 
spin, as given by equation (\ref{eq:e_td}). However, the radiative efficiency is not the same for all regimes of the accretion flow. 
As well as the thin disc and the ADAF case,
there are also AGNs accreting above the Eddington accretion rate. Such objects are generally understood to be
advection dominated and to have optically thick flows \citep{abramowicz88}.

For the ADAF regime we use the expressions for bolometric luminosity from \cite{mahadevan97}. There are two cases
within this regime. For lower accretion rate ADAFs ($\dot{m} < \dot{m}_{\mathrm{crit, visc}}$), heating of the electrons is 
dominated by viscous heating, whereas for higher accretion rate ADAFs ($\dot{m}_{\mathrm{crit, visc}} < \dot{m} < \dot{m}_{\mathrm{crit,ADAF}}$), 
the ion-electron heating dominates the heating of the electrons. In the super-Eddington regime, the radiative efficiency 
is lower than the corresponding thin disc radiative efficiency, and so a super-Eddington luminosity suppression is
introduced \citep{shakurasunyaev73}. This expression includes a free parameter, $\eta_{\mathrm{Edd}}$,
the value for which is given in Table \ref{tab:free_params}, 

Hence, the bolometric luminosities in the model are given by the following expressions\footnote{Note that the coefficients of the ADAF luminosities are 
derived in \cite{mahadevan97} and not free parameters.}. For the low accretion rate ADAF regime, where $\dot{m} < \dot{m}_{\mathrm{crit, visc}}$:

\begin{equation}
 L_{\mathrm{bol}} = 0.0002 \epsilon_{\mathrm{TD}}\dot{M}c^{2} \Big(\frac{\delta_{\mathrm{ADAF}}}{0.0005} \Big) \Big(\frac{1-\beta}{0.5} \Big) \Big( \frac{6}{\hat{r}_{\mathrm{lso}}} \Big). \label{eq:low_adaf}
\end{equation}

\noindent
For the higher accretion rate ADAF regime, where $\dot{m}_{\mathrm{crit, visc}} < \dot{m} < \dot{m}_{\mathrm{crit, ADAF}}$, we have:

\begin{equation}
 L_{\mathrm{bol}} = 0.2 \epsilon_{\mathrm{TD}}\dot{M}c^{2} \Big( \frac{\dot{m}}{\alpha_{\mathrm{ADAF}}^2} \Big) \Big(\frac{\beta}{0.5} \Big) \Big( \frac{6}{\hat{r}_{\mathrm{lso}}}\Big). \label{eq:high_adaf}
\end{equation}

\noindent
For the thin disc regime, where $\dot{m}_{\mathrm{crit, ADAF}} < \dot{m} < \eta_{\mathrm{Edd}}$, $L_{\mathrm{bol}}=L_{\mathrm{bol, TD}}$. Finally, for the super-Eddington regime, where $\dot{m} > \eta_{\mathrm{Edd}}$, we have:

\begin{equation}
 L_{\mathrm{bol}} = \eta_{\mathrm{Edd}}(1+\mathrm{ln}(\dot{m}/ \eta_{\mathrm{Edd}})) \ledd. \label{eq:lbol_se}
\end{equation}

\noindent
The value of $eta_{\mathrm{Edd}}$ adopted gives a similar luminosity at a given mass accretion 
rate in the super-Eddington regime to the model of \cite{watarai00} who model super-Eddington sources as advection dominated slim discs. 

In the above, $\alpha_{\mathrm{ADAF}}$ is the viscosity parameter in the ADAF regime (the value is given in Table \ref{tab:free_params}).
$\delta_{\mathrm{ADAF}}$ is the fraction of viscous energy transferred to the electrons (the value is given in Table \ref{tab:free_params}).
The current consensus for the value of $\delta_{\mathrm{ADAF}}$ is
a value between 0.1 and 0.5, \citep[c.f.][]{yn14}.
Therefore, for this study we adopt a value $\delta_{\mathrm{ADAF}}=0.2$, more in line with observational \citep{yuan03,liu13}
and theoretical \citep{sharma07} constraints, as opposed to the value of $\delta_{\mathrm{ADAF}}=2000^{-1}$ 
adopted in \cite{fani12}. Changing the value of $\delta_{\mathrm{ADAF}}$ makes 
no discernible difference to the  luminosity functions shown in this paper.
$\beta$ is the ratio of gas pressure to total pressure
(total pressure being the sum of gas pressure and magnetic pressure).
Following \cite{fani12}, 
we use the relation $\beta = 1 - \alpha_{\mathrm{ADAF}}/0.55$, which is based on MHD simulations in \cite{hawleygammiebalbus95}.

The boundary between the two ADAF regimes is:

\begin{equation}
 \dot{m}_{\mathrm{crit, visc}} = 0.001 \, \Big( \frac{\delta_{\mathrm{ADAF}}}{0.0005} \Big) \, \Big( \frac{1-\beta}{\beta} \Big) \, \alpha_{\mathrm{ADAF}}^2,
\end{equation}

\noindent
which is a value chosen so that $L_{\mathrm{bol}}$ is continuous in the ADAF regime.
The boundary between the ADAF and thin disc regimes is assumed to be $\dot{m}_{\mathrm{crit, ADAF}} = 0.01$ \citep{yn14}.
$f_{\mathrm{q}}$ and $\eta_{\mathrm{Edd}}$ are free parameters that we calibrate on observed AGN luminosity functions, 
as described in Section \ref{sec:bol_lf_func}.

\subsection{Converting from bolometric to optical and X-ray AGN luminosities}
\label{sec:bol_corr}

\begin{figure}
\centering
\includegraphics[width=\linewidth]{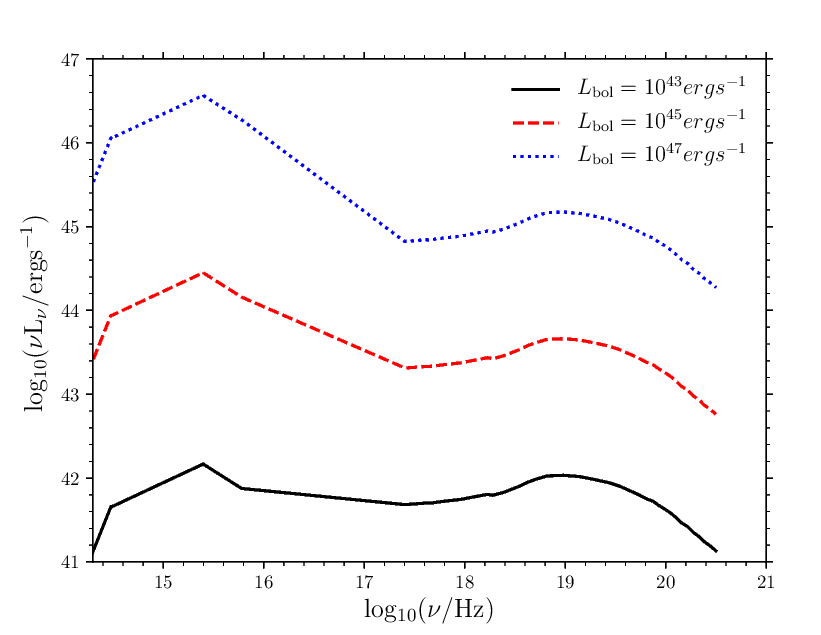}
\caption{The \protect\cite{marconi04} SED used for calculating luminosities in different wavebands in this work. Shown is the SED for $\lbol = 10^{43} \mathrm{ergs^{-1}}$ (black solid line), $\lbol = 10^{45} \mathrm{ergs^{-1}}$ (red dashed line) and for $\lbol = 10^{47} \mathrm{ergs^{-1}}$ (blue dotted line). }
\label{fig:Marconi_SED_example} 
\end{figure}

To convert from AGN bolometric luminosity to luminosities in other wavebands 
we use bolometric corrections derived from the empirical AGN SED template in \cite{marconi04}. We show this SED for three different luminosities in Figure \ref{fig:Marconi_SED_example}. 
The rest-frame bolometric corrections calculated from this SED are\footnote{Note that equations (\ref{eq:bol_corr_hx}) and (\ref{eq:bol_corr_sx}) are corrected from \cite{fani12} equation (10).}:

\begin{equation}
 \mathrm{log}_{10} (L_{\mathrm{HX}} / \lbol) = -1.54 - 0.24 \mathcal{L} - 0.012 \mathcal{L}^{2} + 0.0015 \mathcal{L}^{3}, \label{eq:bol_corr_hx}
\end{equation}

\begin{equation}
 \mathrm{log}_{10} (L_{\mathrm{SX}} / \lbol) = -1.65 - 0.22 \mathcal{L} - 0.012 \mathcal{L}^{2} + 0.0015 \mathcal{L}^{3},\label{eq:bol_corr_sx}
\end{equation}

\begin{equation}
 \mathrm{log}_{10} (\nu_{B} L_{\nu B} / \lbol) = -0.80 + 0.067 \mathcal{L} - 0.017 \mathcal{L}^{2} + 0.0023 \mathcal{L}^{3},
\end{equation}

\noindent
where $\mathcal{L} = \mathrm{log}_{10}(\lbol / 10^{12}L_{\odot})$, $L_{\mathrm{HX}}$ is the hard X-ray (2-10 keV) luminosity, $L_{\mathrm{SX}}$ is the soft X-ray (0.5-2 keV) luminosity, $\nu_{B} = c/ 4400 \angstrom$ is the frequency 
of the centre of the B-band, and $L_{\nu B}$ is the luminosity per unit frequency in the B-band.

To calculate B-band magnitudes we use the expression\footnote{Note that equation (\ref{eq:m_b_ab}) is different to \cite{fani12} equation (13).}:

\begin{equation}
 M_{B, AB} = -11.33 - 2.5 \mathrm{log}_{10} \Big (\frac{\nu_{B} L_{\nu B}}{10^{40} \mathrm{ergs^{-1}}} \Big), \label{eq:m_b_ab}
\end{equation}

\noindent
for magnitudes in the AB system, from the definition of AB magnitudes \citep{okegunn83}. 
Using the \cite{marconi04} SED template, we convert from 
rest-frame B-band magnitudes to rest-frame 1500$\angstrom$ band magnitudes using a relation similar to 
equation (\ref{eq:m1500_m1450}) to give:

\begin{equation}
 M_{1500, AB} = M_{B, AB} + 0.514.
\end{equation}

The \cite{marconi04} SED is based on observations of quasars, with the UV part of the SED based on observations at $L_{UV} \sim 10^{42.5-47} \mathrm{ergs^{-1}}$ and the X-ray part of the SED based on observations at $L_{HX} \sim 10^{41-44} \mathrm{ergs^{-1}}$. Therefore, this SED is likely to be most appropriate for AGN in the thin disc and super-Eddington  
regime. For $z>6$ and for the luminosities that we are considering, the AGN are in the thin disc
or super-Eddington regime, so this SED is appropriate, although
in future work we plan to include a wider variety of SEDs for AGN in different
accretion regimes.

\subsection{AGN obscuration and visible fractions}
\label{sec:unobsc_frac}

AGN are understood to be surrounded by a dusty torus, which causes some of the radiation to be absorbed
along some sightlines,
and re-emitted at longer wavelengths.
For simplicity, we assume that at a given wavelength, AGN are either completely obscured or 
completely unobscured.
The effect of obscuration can therefore be expressed as a visible fraction,
which is the fraction of objects that are unobscured in a certain waveband at a given luminosity and redshift.

The fraction of obscured objects in the hard 
X-ray band is thought to be small, so for this work we assume that there is no obscuration at 
hard X-ray wavelengths.
There is a population of so-called `Compton-thick' AGNs for which the column density of neutral hydrogen exceeds $N_{H} \approx 1.5 \times 10^{24} \mathrm{cm}^{-2}$, which is the unit optical
depth corresponding to the Thomson cross section.
Such objects are difficult to detect, even at hard X-ray wavelengths. The number of such objects is thought to be
small, so we ignore their contribution for this work. 

We calculate the visible fractions in the soft X-ray and optical bands using one of three observationally determined 
empirical relations from the literature, and also two more introduced in this work.

\begin{enumerate}

\item The visible fraction of \cite{hasinger08} is:

\begin{equation}
 f_{\mathrm{vis}} = 1 + 0.281 \big[ \mathrm{log}_{10} \Big( \frac{L_{\mathrm{HX}}}{10^{43.75} \mathrm{ergs^{-1}}} \Big) \big] - A(z),
\end{equation}

\noindent
where 

\begin{equation}
 A(z) = 0.279(1+z)^{0.62}.
\end{equation}

\noindent
$L_{\mathrm{HX}}$ is the hard X-ray luminosity in the observer frame and $z$ is the redshift\footnote{This empirical model and others we use from observational studies were derived using a slightly different 
cosmology from the one used in the P-Millennium, for simplicity we ignore the effect of this here.}. The redshift dependence of the visible fraction  
in this model saturates at $z \geq 2.06$ and the visible fraction is not allowed to have values below 0 or above 1.
Because the observational data on which this obscuration model is based only extend to $z=2$, we extrapolate the model to $z>2$ using $L_{\mathrm{HX}}$ as the rest-frame
hard X-ray band at $z=2$, i.e. 6-30 keV. For this obscuration model, if an object is obscured at soft X-ray wavelengths,
then it is also assumed to be obscured at optical/UV wavelengths.

\item \cite{hopkinsrh07} derive a visible fraction of the form:

\begin{equation}
 f_{\mathrm{vis}} = f_{46} \Big( \frac{\lbol}{10^{46} \rm{ergs^{-1}}}\Big)^{\beta}, \label{eq:hop_vis_frac}
\end{equation}

\noindent
where $f_{46}$ and $\beta$ are constants for each band. For the B-band, [$f_{46}$, $\beta$]
are [0.260, 0.082] and for the soft X-ray band, [$f_{46}$, $\beta$] are [0.609, 0.063]. 
This model does not require a high redshift extrapolation, as it depends only on bolometric luminosity.

\item \cite{aird15} observationally determine a visible fraction for soft X-rays of the form:

\begin{equation}
 f_{\mathrm{vis}} = \frac{\phi_{\mathrm{unabs}}}{\phi_{\mathrm{unabs}} + \phi_{\mathrm{abs}}}, \label{eq:aird_1}
\end{equation}

\noindent
where $\phi_{\mathrm{unabs}}$, the number density of unabsorbed sources, and $\phi_{\mathrm{abs}}$, the number density of absorbed sources, are given by:

\begin{equation}
 \phi = \frac{K}{(\frac{L_{\mathrm{HX}}}{L_{\star}})^{\gamma_{1}} + (\frac{L_{\mathrm{HX}}}{L_{\star}})^{\gamma_{2}}}, \label{eq:aird_2}
\end{equation}

\noindent
where the constants for both cases are given in Table \ref{tab:aird_params}.
As for the \cite{hasinger08} obscuration model, if the object is obscured at soft X-ray wavelengths,
then we assume that it is also obscured at optical/UV wavelengths. For this obscuration model, we extrapolate to high redshift 
such that for $z>3$, the $L_{\mathrm{HX}}$ hard X-ray band is the rest-frame band for $z=3$.

\item We also use visible fractions that are modified versions of \cite{hopkinsrh07}. These visible fractions
also depend solely on $\lbol$, but with different coefficients. These coefficients were derived by 
constructing a bolometric luminosity function from the luminosity functions at optical, UV, and X-ray wavelengths. We used the \cite{marconi04} bolometric corrections and selected coefficients for the
visible fraction so as to create a resultant bolometric luminosity function with the scatter between points minimised. 
This is described in Appendix \ref{app:visible_fraction}.
The first of these new obscuration relations, the `low-z modified Hopkins', 
(LZMH) visible fraction for rest-frame $1500\angstrom$ has the form:

\begin{equation}
 f_{\mathrm{vis, LZMH}} = 0.15 \Big( \frac{\lbol}{10^{46} \rm{ergs^{-1}}}\Big)^{-0.1}, \label{eq:LZMH_opt}
\end{equation}

\noindent
and for the soft X-ray band it has the form:

\begin{equation}
 f_{\mathrm{SX, LZMH}} = 0.4 \Big( \frac{\lbol}{10^{46} \rm{ergs^{-1}}}\Big)^{0.1}. \label{eq:LZMH_SX}
\end{equation}

\item The second of these modified Hopkins visible fractions, the `$z=6$ modified Hopkins' 
(Z6MH) visible fraction was derived by fitting the \galform $z=6$ luminosity functions at $1500 \angstrom$ and in the soft X-ray band to the observational estimates. This visible fraction is:

\begin{equation}
 f_{\mathrm{vis, Z6MH}} = 0.04, \label{eq:Z6MH}
\end{equation}

\noindent 
for both rest-frame $1500 \angstrom$ and soft X-rays.

\begin{table*}
\caption{The parameters that correspond to the best fit visible fraction from \protect\cite{aird15} where 
$\zeta = \mathrm{log}(1+z)$. These parameter values have been obtained by private communication.
See equations (\ref{eq:aird_1}) and (\ref{eq:aird_2}).}
\begin{tabular}{ |c|c|c| }
\hline
& absorbed & unabsorbed \\ \hline
log(K $\rm{/ Mpc^{-3}}$) & $-4.48 + 3.38 \zeta - 7.29 \zeta^2$ & $-5.21 + 3.21 \zeta - 5.17 \zeta^2$ \\ 
$\rm{log} (L_{\star}/ \rm{ergs^{-1}})$ & $43.06 + 3.24 \zeta - 1.59 \zeta^2 + 0.43 \zeta^3$ & $43.80 - 0.57 \zeta + 9.70 \zeta^2 - 11.23 \zeta^3$ \\ 
$\rm{log} \gamma_{1}$ & $-0.28 - 0.67 \zeta$ & $-0.44 -1.25 \zeta$ \\ 
$\gamma_{2}$ & 2.33 & 2.32 \\ 
$\beta_{CT}$ & 0.34 & 0.34 \\
\hline
\end{tabular}
\label{tab:aird_params}
\end{table*}

\end{enumerate}

\subsection{Calculating model AGN luminosity functions}
\label{sec:time_av}

Typically when one constructs a luminosity function from a simulation, only the AGN that are switched on at 
each snapshot are included. However, if one does this, rarer objects with higher luminosities but which are only active
for a short time are not sampled well. To probe the luminosity function for such objects, we average over a time window, $\Delta t_{\mathrm{window}}$.
The time window should not be too large, as then we may miss the effect of multiple
starbursts within the time window,
because the simulation only outputs information on the most recent starburst. 
We select a time window for which the luminosity function using the time average method is converged to the luminosity function using only the AGN switched on at the snapshots. For the predictions here we set $\Delta t_{\mathrm{window}}=t_{\mathrm{snapshot}}/10$, where 
$t_{\mathrm{snapshot}}$ is the age of the Universe at that redshift.

Each object is assigned a weight, $w$, given by:

\begin{equation}
 w = t_{Q} / \Delta t_{\mathrm{window}},
\end{equation}

\noindent
where $t_{Q}=f_{\mathrm{q}}t_{\mathrm{bulge}}$ is the lifetime of the most recent quasar episode occurring within the time interval $\Delta t_{window}$ as in Section \ref{sec:lbol}.
This weight is then applied to the number densities counting all AGN occurring within the time interval $\Delta t_{window}$ which then allows us to include higher luminosity 
events at lower number densities in the luminosity function.
We show the effect of changing the value of $\Delta t_{window}$, as well as the effect of simply using snapshot quantities 
on the predicted luminosity functions in Appendix \ref{app:time_averaging}.


\section{SMBH Masses, accretion rates and spins}
\label{sec:preliminaries}

\begin{figure*}
\centering
\includegraphics[width=.8\linewidth]{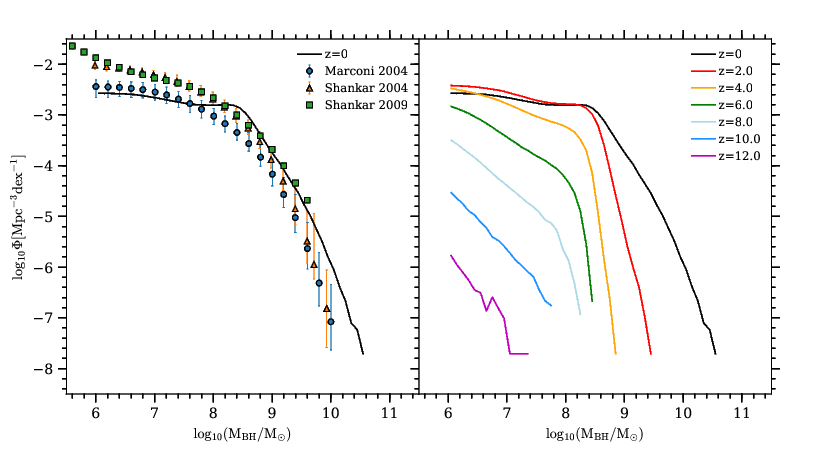}
\caption{The black hole mass function. \emph{Left panel}: the predicted black hole mass function at $z=0$ compared to 
observational estimates by \protect\cite{marconi04,shankar04,shankar09}.
\emph{Right panel}: the evolution of the black hole mass function over the range 0 < z < 12.}
\label{fig:BHMF_evolution} 
\end{figure*}

We start by showing some basic predictions from the new model for SMBH masses, accretion rates and spins.

\subsection{Black hole masses}

In the left panel of Figure \ref{fig:BHMF_evolution} we show the black hole mass function at $z=0$ predicted by our model 
compared to observational estimates.
The observations use indirect methods to estimate the black hole mass function, because of the lack
of a large sample of galaxies with dynamically measured black hole masses.
In \cite{marconi04} and \cite{shankar04,shankar09} galaxy luminosity/velocity dispersion functions are 
combined with relations between black hole mass and host galaxy properties to estimate black hole mass functions. 
The predictions of the model fit well to the observational estimates within the observational errors, especially given that there will also 
be uncertainties on the black hole mass measurements and given the discrepancies between the observational estimates.
The former means the predictions could still be consistent with observations at the high mass end ($\mbh \geq 10^{9} M_{\odot}$).

The evolution of the black hole mass function for $0 < z < 12$ is shown in the right panel of Figure \ref{fig:BHMF_evolution}.
Most of the SMBH mass is formed by $z \sim 2$, as the mass density of black holes is dominated by objects around 
the knee of the black hole mass function, and this
knee is in place by $z \sim 2$. The dominant fuelling mechanism for growing the black hole mass density across all redshifts 
is gas accretion in starbursts triggered by disc instabilities, and disc instabilities
play an important role in shaping the black hole mass function for $\mbh < 10^{8} M_{\odot}$. However, SMBH mergers
are more important for determining the shape of the black hole mass function for $\mbh > 10^{8} M_{\odot}$, as they are
the mechanism by which the largest SMBHs are formed. AGN feedback also plays an important role in shaping 
the black hole mass function at this high mass end, by suppressing gas cooling and so slowing down the rate 
at which the SMBHs grow by cold gas accretion.

\begin{figure}
\centering
\includegraphics[width=\linewidth]{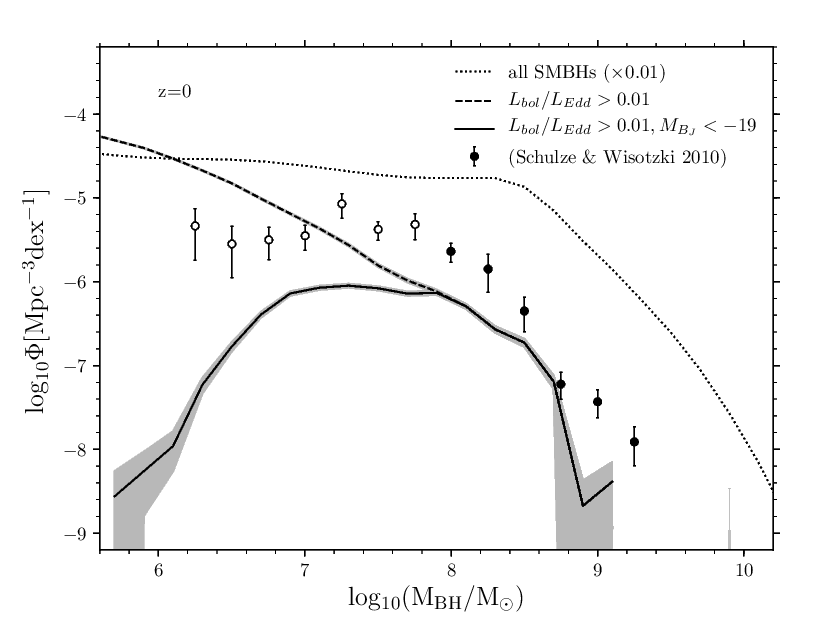}
\caption{The active black hole mass function (solid line) at $z=0$, compared to observational estimates from \protect\cite{schulze10}. We show 
predictions where active SMBHs are defined as AGN brighter than a threshold Eddington ratio ($L_{\mathrm{bol}}/\ledd >0.01$), using the LZMH visible fraction (c.f. Section \ref{sec:unobsc_frac}) (dashed line), and predictions also brighter than a threshold AGN absolute magnitude ($M_{\mathrm{B_{J}}}<-19$) (solid line). 
This is for appropriate comparison with the active black hole mass function 
in \protect\cite{schulze10}, where the open circles are the data points that suffer from incompleteness, while the filled circles are the data points that do not. 
We also show the total black hole mass function (dotted line) with the number density divided by 100, for comparison.}
\label{fig:BHMF_active} 
\end{figure}

In Figure \ref{fig:BHMF_active}, we show the `active' black hole function at $z=0$ compared to observational estimates from \cite{schulze10}. In this observational estimate, active SMBHs are defined as AGN radiating above a certain Eddington ratio ($L_{\mathrm{bol}}/\ledd > 0.01$). The flux limit in the observations results in the observational sample being incomplete for $M_{B_{J}} > -19$. The observational sample also only includes type 1 (unobscured) AGN. Therefore, we apply these selections to the model predictions, using the LZMH visible fraction, to compare with this observational estimate of the active black hole mass function. We also present predictions where the selection on $M_{B_{J}}$ has not been applied. The effect of the selection on $M_{B_{J}}$ can be seen at the low mass end ($\mbh < 10^8 M_{\odot}$), where the dashed and solid lines diverge. While the model is in reasonable agreement with the observations at $\mbh \sim 10^{8.5} M_{\odot}$, the model generally underpredicts the active black hole mass function, although the model does reproduce the overall shape of the shape of the observational active black hole mass function. We found similar results when comparing with other studies, such as those from SDSS \citep[e.g.][]{vestergaard09}.

\begin{figure*}
\centering
\includegraphics[width=\linewidth]{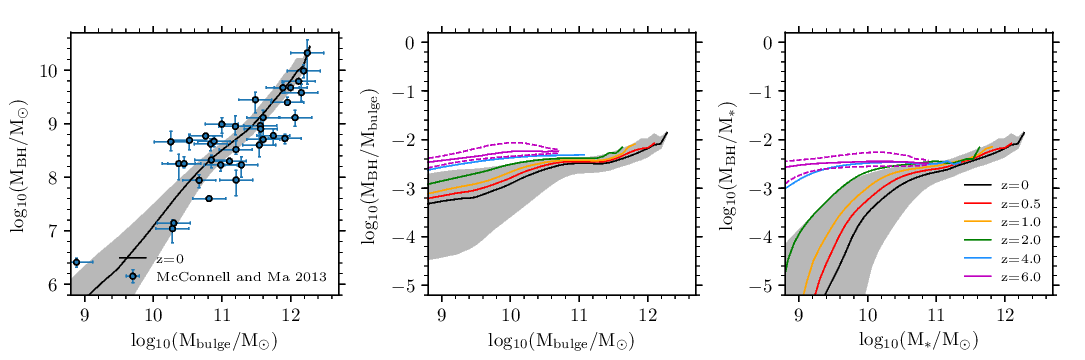}
\caption{\emph{Left panel}: the predicted SMBH mass versus bulge stellar mass relation at $z=0$ compared to observational data from \protect\cite{mcconnellma13}. 
The line represents the median of the predicted SMBH mass in bins of bulge mass and the shading denotes the 10-90 percentiles of 
the predicted distribution. \emph{Middle panel}: the evolution of the median of the ratio of SMBH mass to bulge mass versus bulge mass relation with redshift
for $z$ = 0, 0.5, 1, 2, 4, 6. As in the left panel, 
the grey shaded band is the 10-90 percentiles of the distribution for $z=0$ and the purple dashed
lines are the 10-90 percentiles of the distribution for $z=6$. \emph{Right panel}: the evolution of the median 
of the ratio of SMBH mass to galaxy stellar mass versus galaxy stellar mass relation, with the lines representing the same redshifts as the middle panel
as indicated by the legend.}
\label{fig:SMBH_Mbulge_Mstar_relation} 
\end{figure*}

Figure \ref{fig:SMBH_Mbulge_Mstar_relation} shows the relation between SMBH mass and bulge or total stellar mass.
In the left panel of Figure \ref{fig:SMBH_Mbulge_Mstar_relation} we show the predicted SMBH mass versus bulge mass relation compared to 
observational data from \cite{mcconnellma13}.
The predictions follow the observations well, with the scatter decreasing towards higher masses.
BH-BH mergers contribute towards this decrease in scatter, as seen in \cite{jahnkemaccio11},
although they are not the only contributing mechanism, with AGN feedback also affecting the scatter at the high
mass end.

In the middle panel of Figure \ref{fig:SMBH_Mbulge_Mstar_relation}, we show the evolution of the ratio of SMBH mass to bulge mass ($\mbh/M_{\mathrm{bulge}}$) versus bulge stellar mass for $0 < z < 6$, showing
the scatter of the distribution for $z=0$ and $z=6$. 
As we go to higher redshift, the ratio $\mbh/M_{\mathrm{bulge}}$ increases, as also seen in observations
\citep[e.g.][]{peng06}. The ratio $\mbh/M_{\mathrm{bulge}}$ reflects the mechanism by which these two galaxy components
form. At higher redshift, bulges grow mainly by starbursts, which also feeds the growth of SMBHs and so the distribution of the ratio
$\mbh/M_{\mathrm{bulge}}$ peaks at $f_{\mathrm{BH}}$ (the fraction of the mass of stars formed in a starburst accreted onto a black hole), with some scatter caused by mergers. At lower redshift the ratio 
$\mbh/M_{\mathrm{bulge}}$ decreases, as galaxy mergers cause bulges to form from discs, but without growing the SMBHs.
We also note how the scatter of the relation is lower at $z=6$ than at $z=0$ for all masses - 
by $z=0$ galaxies have had more varied formation histories compared to the $z=6$ population.

In the right panel of Figure \ref{fig:SMBH_Mbulge_Mstar_relation} we show the evolution of the ratio of the SMBH mass to the galaxy stellar mass ($\mbh / M_{\star}$) versus galaxy stellar mass for 
the redshift range $0 < z < 6$. Galaxies of larger stellar mass and the largest SMBHs form at late times, 
and at lower masses ($M_{\star} < 10^{11} M_{\odot}$), $\mbh / M_{\star}$ is smaller at later times.
At lower masses, the ratio $\mbh / M_{\star}$
decreases with time because the fraction of the stellar mass that is in the bulge decreases.
This evolution slows down at $z<1$.
At higher masses ($M_{\star} > 10^{11} M_{\odot}$), the stellar mass and SMBH mass stay on the same relation independent of redshift.
It is in this regime that the AGN feedback is operational: in our model we use the AGN feedback prescription of 
\cite{bower06} in which AGN feedback is only active where the hot gas halo is undergoing `quasistatic' (slow) cooling.
This has the effect that AGN feedback is only active for haloes of mass above $\sim 10^{12} M_{\odot}$.
The relation between SMBH mass and stellar mass at this high mass end is caused by both AGN feedback and mergers,
with neither mechanism dominant in establishing this relation.

\begin{figure*}
\centering
\includegraphics[width=.8\linewidth]{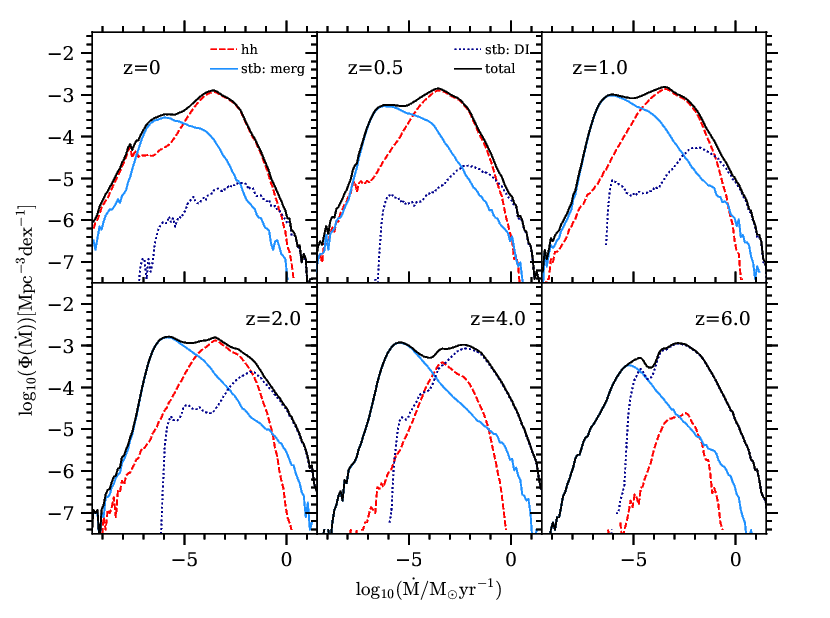}
\caption{The distribution of black hole mass accretion rates for different redshifts (black solid line) split by contributions
from hot halo mode (red dashed line), starbursts triggered by mergers (light blue solid line) and starbursts triggered by disc instabilities
(dark blue dotted line). We have selected all black holes residing in galaxies of stellar mass, $M_{\star}>10^6 M_{\odot}$,
which is above the completeness limit of the simulation.}
\label{fig:accretion_rate_distribution} 
\end{figure*}

\subsection{Black hole accretion rates}

In Figure \ref{fig:accretion_rate_distribution} we show the black hole mass accretion rate distribution, showing its evolution with redshift 
and split by fuelling modes: the hot halo mode, starbursts triggered by mergers and starbursts triggered by disc instabilities 
(see Section \ref{sec:model}). The hot halo mode becomes more dominant at later times, because the hot halo 
mode requires long cooling times, and hence it occurs for massive haloes, and because dark matter 
haloes grow hierarchically, these large haloes only form at later times. 
The contribution from starbursts triggered by galaxy mergers peaks at $z \approx 2$. 
Starbursts triggered by mergers peak at a low mass accretion rate, 
as seen in Figure \ref{fig:accretion_rate_distribution}, albeit with a tail that extends to high $\dot{M}$. 
The peak at $\dot{M} \sim 10^{-6} M_{\odot}$/yr is mostly due to minor mergers with mass ratios $0.05 < M_{2}/M_{1} < 0.3$ (mergers with mass
ratios in this range cause about three quarters of the merger triggered starbursts at this mass accretion rate)\footnote{Note that a mass ratio of 0.05 is assumed to be the lower threshold for starburst triggering in galaxy mergers \citep{lacey16}}. 
The contribution from starbursts
triggered by disc instabilities increases as the redshift increases. Starbursts triggered by mergers 
typically have lower $\dot{M}$ values than starbursts triggered by disc instabilities.
There are two reasons for this.
Firstly, the average stellar mass formed by bursts triggered by disc instabilities is higher than for 
bursts triggered by mergers, and this occurs because the average cold gas mass is higher for galaxies
in which bursts triggered by disc instabilities occur. 
Secondly, the average bulge dynamical timescale for starbursts triggered by disc instabilities is smaller than for those 
triggered by mergers due to the average bulge size being smaller for starbursts triggered by
disc instabilities. The combination of these effects accounts for the lack of 
starbursts triggered by disc instabilities at the very lowest 
$\dot{M}$ values. The galaxies that host such starburst episodes would be below the 
mass at which the simulation is complete.

\begin{figure*}
\centering
\includegraphics[width=.8\linewidth]{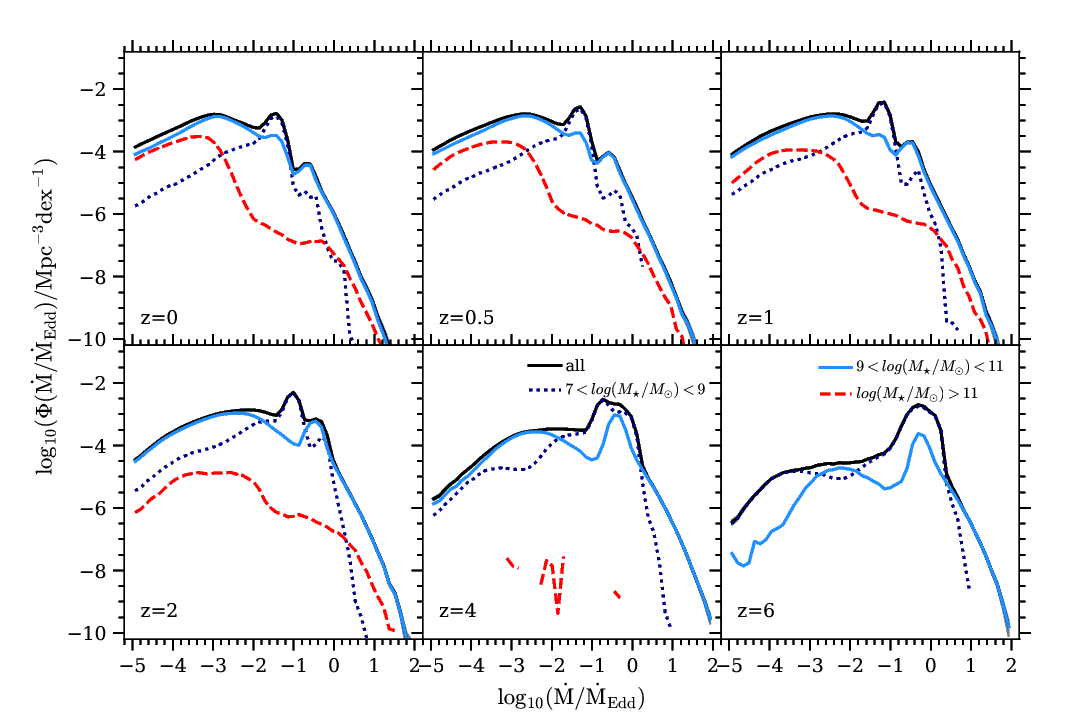}
\caption{The distribution of Eddington ratio in terms of mass accretion rate, $\dot{M}/ \dot{M}_{\mathrm{Edd}}$, evolving with redshift. 
Shown are all objects with stellar mass, $M_{\star} > 10^{7} M_{\odot}$ (black solid line), objects with stellar mass $10^{7} M_{\odot} < M_{\star} < 10^{9} M_{\odot}$ (dark blue dotted line), objects with stellar mass $10^{9} M_{\odot} < M_{\star} < 10^{11} M_{\odot}$ (light blue solid line) and objects with stellar mass $M_{\star} > 10^{11} M_{\odot}$ (red dashed line).}
\label{fig:MMEdd_distribution} 
\end{figure*}

In Figure \ref{fig:MMEdd_distribution} we show the evolution of the distribution of Eddington normalised mass accretion rate $\dot{M}/\dot{M}_{\mathrm{Edd}}$. We also show the predictions in different stellar mass ranges. Looking at the total distribution ($M_{\star} > 10^{7} M_{\odot}$), for increasing redshift, the distribution shifts to somewhat higher values. This is seen as the number of objects with log$(\dot{M}/ \dot{M}_{\mathrm{Edd}}) < -2$ decreasing with increasing redshift, a peak at log$(\dot{M}/ \dot{M}_{\mathrm{Edd}}) \sim -1$ building up with increasing redshift and the number of objects with log$(\dot{M}/ \dot{M}_{\mathrm{Edd}}) > 0$ increasing with increasing redshift. The different bins of stellar mass have different distributions of $\dot{M}/\dot{M}_{\mathrm{Edd}}$, and evolve differently. At $z=0$, the lowest bin in stellar mass ($10^{7} M_{\odot} < M_{\star} < 10^{9} M_{\odot}$) shows a broad distribution around a peak at log$(\dot{M}/ \dot{M}_{\mathrm{Edd}}) \approx -1.5$, the middle bin in stellar mass ($10^{9} M_{\odot} < M_{\star} < 10^{11} M_{\odot}$) also shows a broad distribution, but with a peak at log$(\dot{M}/ \dot{M}_{\mathrm{Edd}}) \approx -3$ and also has features at log$(\dot{M}/ \dot{M}_{\mathrm{Edd}}) \approx -1.5$ and log$(\dot{M}/ \dot{M}_{\mathrm{Edd}}) \approx -0.5$. The distribution in the highest stellar mass bin ($M_{\star} > 10^{11} M_{\odot}$) peaks at lower value of log$(\dot{M}/ \dot{M}_{\mathrm{Edd}}) \approx -4$, but has fewer objects at high Eddington ratios than the lower stellar mass bins. The distribution in the highest stellar mass bin peaks at a lower Eddington ratio because this is where the hot halo mode is operational, so SMBHs are typically quiescently accreting.

As redshift increases, the number density at the peak in the $\dot{M}/\dot{M}_{\mathrm{Edd}}$ distribution for the lowest stellar mass bin increases, such that by $z=6$, the peak for the lowest stellar mass bin and the middle stellar mass bin are both at log$(\dot{M}/ \dot{M}_{\mathrm{Edd}}) \approx -0.5$. The number of objects in the highest stellar mass bin decreases strongly at high redshift, so the hot halo mode is much less prevalent at higher redshift, $z>3$.

We also have compared the predicted Eddington luminosity ratio, ($\lbol / \ledd$) distribution at $z=6$, to the observational data compiled in \cite{wu15} Figure 4. The $\lbol / \ledd$ distribution at $z=6$ from \galform has a median and 10-90 percentiles at $4.3^{+4.3}_{-3.0}$ for AGN with $\lbol > 10^{46} \mathrm{ergs^{-1}}$) and $8.6^{+3.5}_{-3.5}$ for AGN with $\lbol > 10^{47} \mathrm{ergs^{-1}}$, whereas the $\lbol / \ledd$ median and 10-90 percentiles in \cite{wu15} is $1.0^{+1.8}_{-0.4}$ for a mixture of samples with $L_{\mathrm{bol}} > 10^{46} ergs^{-1}$. The predicted $\lbol / \ledd$ are somewhat larger than the observational estimate. One possible reason for the different distributions is systematic uncertainties in the black hole mass estimates in the observations. We plan to conduct a more detailed investigation in future work.

\subsection{Black hole spins}

\begin{figure*}
\centering
\includegraphics[width=.8\linewidth]{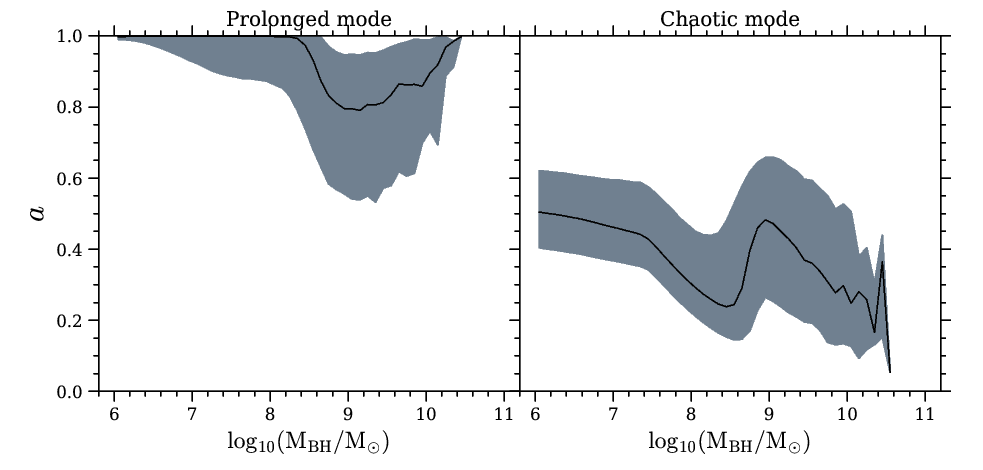}
\caption{The predcited SMBH spin distributions at $z=0$ for prolonged (left panel) and chaotic (right panel) accretion modes. 
The line represents the median value of the magnitude of the spin
for that SMBH mass, and the shading represents the 10-90 percentile range of the distribution.}
\label{fig:spin_vs_mbh} 
\end{figure*}

In Figure \ref{fig:spin_vs_mbh} we show the SMBH spin distribution predicted by the model for both the prolonged and 
chaotic accretion modes. Note that $a$ here represents the magnitude of the spin. 
The low mass end of the spin distribution (6 < $\mathrm{log_{10}} (\mbh / M_{\odot})$ < 8) is dominated by accretion spinup 
whereas the high mass end (8 < $\mathrm{log_{10}} (\mbh / M_{\odot})$ < 10) is
dominated by merger spinup. For prolonged mode accretion, the coherent accretion spinup means that SMBHs quickly reach their 
maximum spin value, giving rise to a population of maximally spinning SMBHs at low mass. At high masses,
the average spin value is lower because of SMBH mergers. This is because even if two maximally spinning SMBHs 
merge, the result is typically a SMBH with a lower spin value because of misalignment between the black hole spins and the orbital angular momentum. For chaotic mode accretion, the accretion direction is constantly 
changing and so the accretion spinup leads to SMBHs with lower median spin values ($a \approx 0.4$), compared to prolonged
accretion. The spin values are not zero in the chaotic mode, as one may be tempted to expect,
because the accretion spinup is more efficient if the accretion disc and SMBH spin are in the same direction compared to 
the case of anti-alignment \citep{kph08}. The mean value of the SMBH spin decreases with increasing black hole mass at this low mass end, 
for chaotic mode accretion as also reported in \cite{kph08}.
At the high mass end, the increase in average spin at $\mbh \sim 10^9 M_{\odot}$ is due to 
spinup by BH mergers. Two slowly spinning SMBHs typically form a higher spin SMBH when they merge,
due to the angular momentum of the orbit between them.

One of the conclusions of \cite{fani11}
was that for chaotic mode accretion, smaller SMBHs will have lower spin values ($\bar{a} \approx 0.15$) whereas larger SMBHs 
will have higher spin values ($\bar{a} \approx 0.7-0.8$). Our new analysis predicts that for chaotic mode accretion SMBHs
will generally have moderate spin values, $\bar{a} \approx 0.4$, yielding radiative accretion efficiencies of $\epsilon \approx 0.075$,  
not too dissimilar from the value of $\epsilon \approx 0.1$ required by the \cite{soltan82} argument. However, the average radiative accretion 
efficiency implied by prolonged mode accretion is $\epsilon \approx 0.4$, in tension with the \cite{soltan82}
argument.

The chaotic mode spin distribution is different
to that in \cite{fani11} because the equations for SMBH spinup by gas accretion have changed from that paper (causing higher spin values at the low
SMBH mass end) and because the directions for the spinup due to SMBH mergers are sampled from the surface of a sphere as opposed to the circumference
of a circle, leading to lower spin values at the high SMBH mass end.

\begin{figure*}
\centering
\includegraphics[width=.8\linewidth]{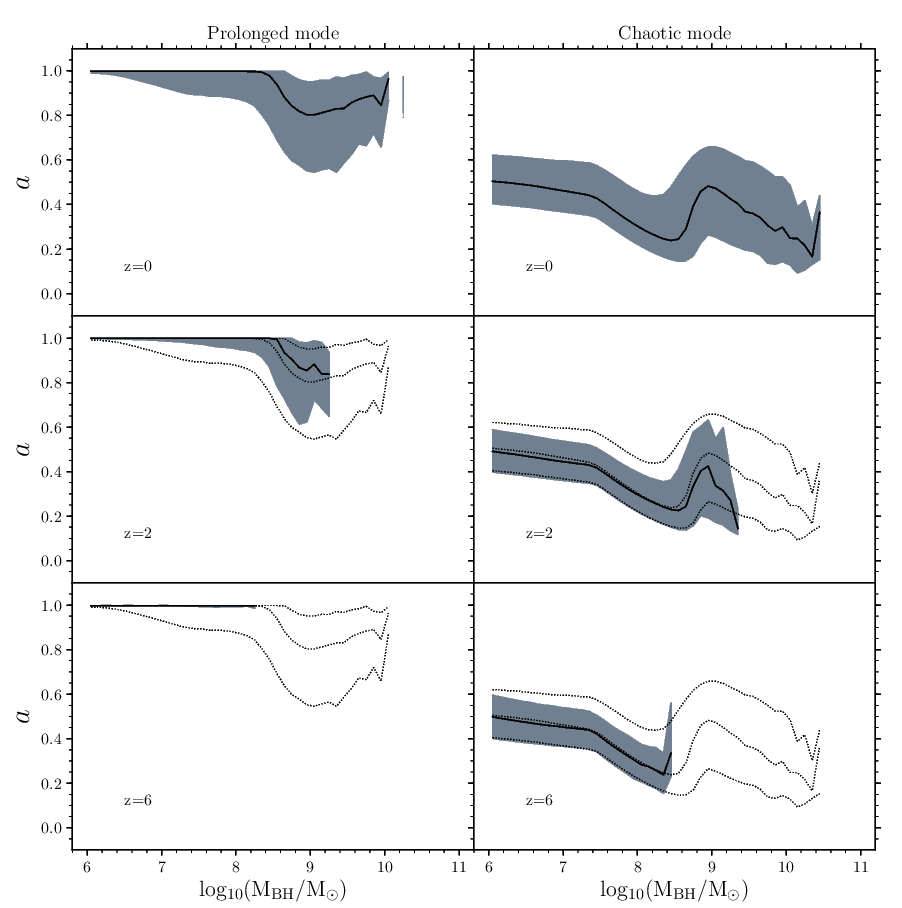}
\caption{The predicted evolution of the SMBH spin distribution for prolonged mode (left panels) and chaotic mode (right panels). 
Results are shown for $z=0,2,6$.
The lines and shading have the same meaning as in the previous figure, with the dotted line representing the median and percentiles for that accretion mode at $z=0$.}
\label{fig:spin_vs_mbh_evolution} 
\end{figure*}

We then show the evolution of the SMBH spin distribution for the prolonged and chaotic modes in Figure 
\ref{fig:spin_vs_mbh_evolution}. 
The black hole spin versus black hole mass relation shows negligible evolution for both modes, 
with the median black hole spin at any black hole mass approximately the same over the range $z=0-6$. 
For both modes the scatter of the distribution decreases with increasing redshift, with the scatter for the 
prolonged mode decreasing much more than the scatter for the chaotic mode. For the prolonged mode, by $z=6$, nearly all of the black holes with $\mbh < 10^{8} M_{\odot}$ have the maximal spin permitted by the model.
Also, there is a lack of high 
mass, $\mbh > 3 \times 10^8 M_{\odot}$, black holes at $z=6$ for both modes. This is due to a low abundance of high mass galaxies at $z=6$.

\begin{figure}
\centering
\includegraphics[width=1\linewidth]{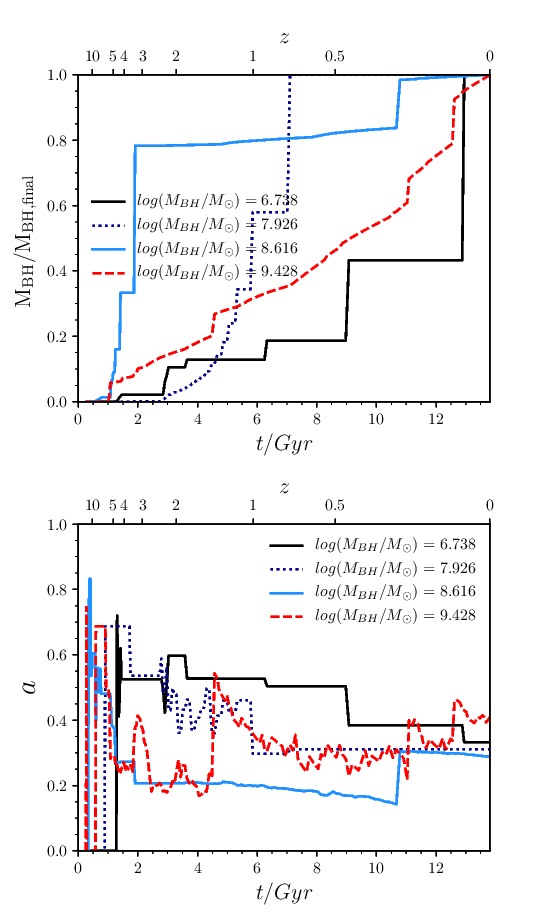}
\caption{\emph{Upper panel:} the evolution of the ratio of SMBH mass to the SMBH at $z=0$ versus time. \emph{Lower panel:} the evolution of SMBH spin versus time. In both panels we show examples of SMBHs with $z=0$ masses of $\mbh = 5.47 \times 10^{6} M_{\odot}$ (black solid line), $\mbh = 8.43 \times 10^{7} M_{\odot}$ (dark blue dotted line), $\mbh = 4.13 \times 10^{8} M_{\odot}$ (light blue solid line), $\mbh = 2.68 \times 10^{9} M_{\odot}$ (red dashed line). The same objects are plotted in both panels.}
\label{fig:TD_ambh_t}
\end{figure}

\begin{figure}
\centering
\includegraphics[width=1\linewidth]{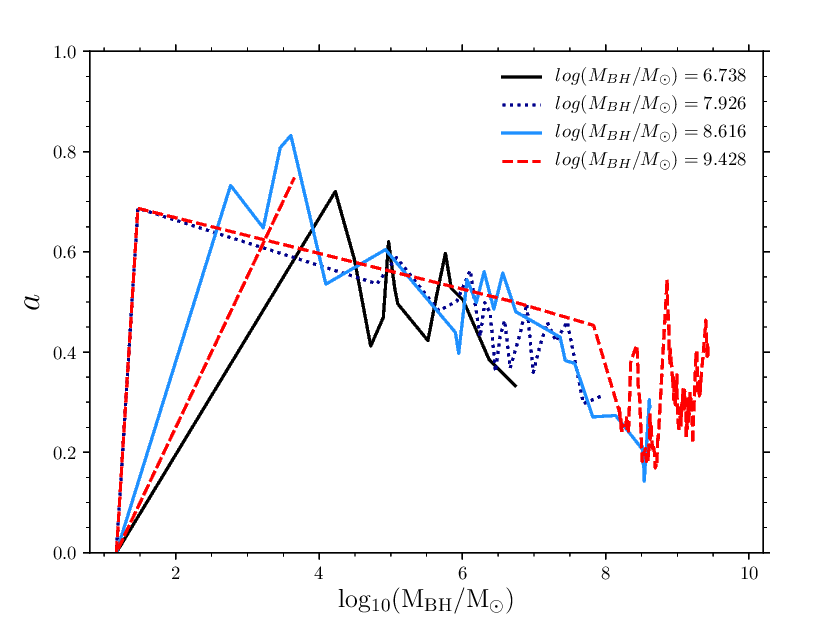}
\caption{The evolution of four different mass SMBHs through the spin versus mass plane. The final SMBH masses at $z=0$ are the same as plotted in Figure \ref{fig:TD_ambh_t}.}
\label{fig:TD_a_mbh}
\end{figure}

We show how typical black holes evolve in the chaotic mode (the standard choice for this analysis) for four different black hole masses in Figures in \ref{fig:TD_ambh_t} and \ref{fig:TD_a_mbh}. When we generate each 
black hole history, we only follow the largest progenitor black hole back in time when two or more black holes merge. In the upper panel of Figure \ref{fig:TD_ambh_t} we show the evolution of the black hole mass through time evolution for these objects, where the time is measured from the Big Bang. Some of the features discussed for the black hole mass function in Figure~\ref{fig:BHMF_evolution} can be seen here, such as how most of the SMBH mass is assembled at early times, and how the very largest black holes build up gradually at late times. It can also be seen how the larger SMBHs generally grow their mass quickest, with smaller SMBHs generally growing later. This is seen in Figure \ref{fig:TD_ambh_t} where the SMBH of mass $\mbh = 5.47 \times 10^{6} M_{\odot}$ reaches $40 \%$ of its final mass at 9 Gyr, whereas the SMBH of mass $\mbh = 8.43 \times 10^{7} M_{\odot}$ reaches $60 \%$ of its final mass at 6 Gyr, and the SMBH of mass $\mbh = 4.13 \times 10^{8} M_{\odot}$ reaches $80 \%$ of its final mass at 2 Gyr. However, the SMBH of mass $\mbh = 2.68 \times 10^{9} M_{\odot}$ grows more gradually.

In the lower panel of Figure \ref{fig:TD_ambh_t} we show the evolution of SMBH spin through time. SMBHs of different masses generally show the same trends as their spin evolves through time. At early times, the black holes are smaller and so the spin values will change dramatically (with $a$ changing between 0 and 0.8) if there is an accretion or merger event, whereas at later times, the spin values do not change as dramatically ($a$ only varies by about 0.1 for each event) with time. The spin values generally converge on a moderate value ($a \approx 0.2 - 0.6$) at late times. 

In Figure \ref{fig:TD_a_mbh}, we show the evolution of the black holes through the spin versus mass plane. First, the black holes are spun up to high spins by mergers at small masses. Then the black holes of different masses generally show a similar evolution through the spin versus black hole mass plane as they evolve from high spins at lower black hole masses to lower spins at higher black hole masses, as they accrete gas by chaotic mode accretion. For the two largest black hole masses, there is an additional feature, as the black hole spin increases at the very highest masses. This is a result of the black holes merging with other black holes following their host galaxies merging.


\subsection{AGN luminosities and black hole/galaxy properties}
\label{sec:L_AGN_galaxy}

\begin{figure*}
\centering
\includegraphics[width=.8\linewidth]{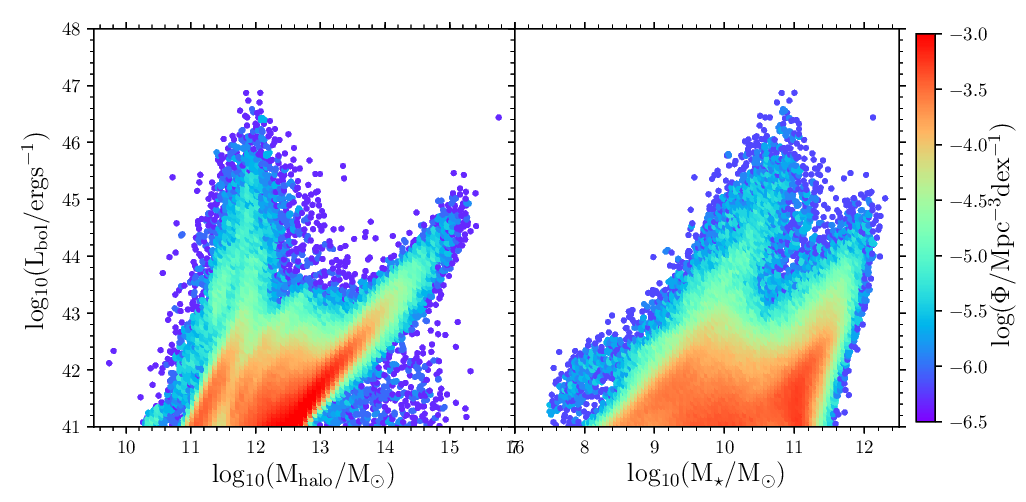}
\caption{\emph{Left panel:} a scatter plot of AGN bolometric luminosity versus halo mass at $z=0$. The points are coloured 
by the density of objects in this plane, 
where red indicates a high density of objects while blue indicates a low density of objects. \emph{Right panel:} as in the left panel but showing bolometric luminosity versus stellar mass.}
\label{fig:Lbol_vs_mstar_mhhalo_cs} 
\end{figure*}

Before comparing the predicted AGN luminosity functions to observational estimates, we first show the dependence of AGN 
luminosities on some different galaxy properties. 

First in the left panel of Figure \ref{fig:Lbol_vs_mstar_mhhalo_cs}, we show the dependence of bolometric luminosity on halo mass, where the points are coloured by the density of points.
Each halo mass can host an AGN up to $L_{\mathrm{bol}} \sim 10^{44} \mathrm{ergs^{-1}}$, with the brightest AGN 
not residing in the largest haloes, but instead 
in haloes of mass $M_{\mathrm{halo}} \sim 10^{12} M_{\odot}$. This is a result of how in the model, AGN activity is inhibited in the largest haloes due to AGN feedback \citep[c.f.][]{fani13}. The overall distribution is bimodal, which is a result of the two primary fuelling modes. 
The AGN at $M_{\mathrm{halo}} \lesssim 10^{12.5} M_{\odot}$ are mostly fuelled by starbursts triggered by disc instabilities, whereas the AGN at $M_{\mathrm{halo}} \gtrsim 10^{12.5} M_{\odot}$ are mostly fuelled by hot halo mode accretion. AGN fuelled by starbursts triggered by mergers make a minor contribution to both parts of this distribution.  
Hot halo mode accretion fuels the objects at the peak of the 2D distribution in this plane seen at $M_{\mathrm{halo}} \approx 10^{13} M_{\odot}$ and $L_{\mathrm{bol}} \approx 10^{42} \mathrm{ergs^{-1}}$. 
The peak of the distribution of objects fuelled by starbursts triggered by disc instabilities is at $M_{\mathrm{halo}} \approx 10^{11.5} M_{\odot}$ and $L_{\mathrm{bol}} \approx 10^{43.5} \mathrm{ergs^{-1}}$, while the peak in the distribution for starbursts triggered by mergers is at $M_{\mathrm{halo}} \approx 10^{11.5} M_{\odot}$ and $L_{\mathrm{bol}} \approx 10^{42} \mathrm{ergs^{-1}}$.

In the right panel of Figure \ref{fig:Lbol_vs_mstar_mhhalo_cs}, we show the dependence of bolometric luminosity on stellar mass. 
There is more of a correlation between bolometric luminosity and stellar mass than between bolometric luminosity 
and halo mass. The brightest AGN in the model do not live in the largest stellar mass galaxies, but rather reside in galaxies 
of $M_{\star} \sim 10^{11} M_{\odot}$.
This distribution also shows a bimodality, where generally the objects at lower masses ($M_{\star} < 3 \times 10^{10} M_{\odot}$) are fuelled by the starburst mode, while objects at higher masses ($M_{\star} > 3 \times 10^{10} M_{\odot}$) are fuelled by the hot halo mode, although there is some overlap between the two. For the starburst mode, the peak of the distribution for starbursts triggered by disc instabilities and the peak of the distribution for starbursts triggered by mergers are both at stellar mass $M_{\star} \approx 3 \times 10^{9} M_{\odot}$. This peak is at $\lbol \approx 10^{43} \mathrm{ergs^{-1}}$ for disc instabilities, whereas for mergers this peak is at $\lbol \approx 10^{42} \mathrm{ergs^{-1}}$. Starbursts triggered by mergers do also occur for galaxies of stellar mass $M_{\star} > 10^{11} M_{\odot}$, whereas starbursts triggered by disc instabilities do not occur for galaxies of this mass.

\begin{figure}
\centering
\includegraphics[width=\linewidth]{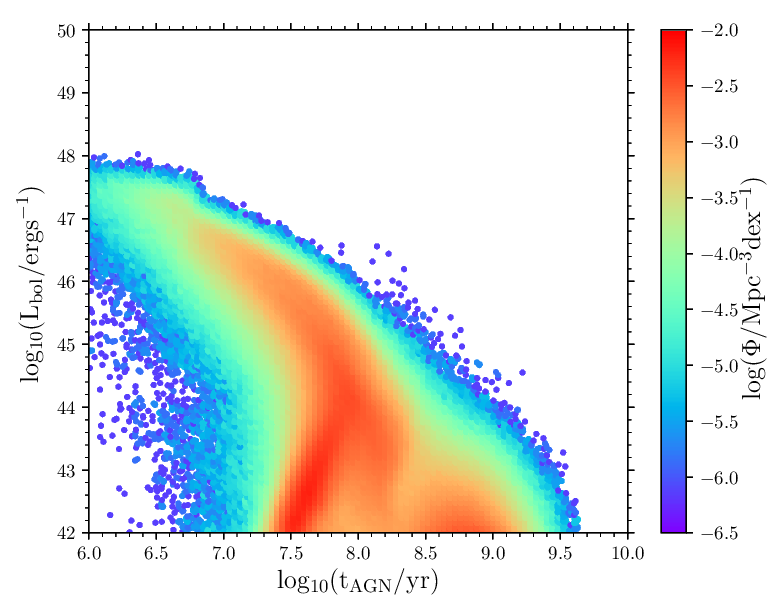}
\caption{As in Figure \ref{fig:Lbol_vs_mstar_mhhalo_cs} but showing the dependence of AGN bolometric luminosity 
on the duration of the AGN episode, for starburst mode fuelled AGN only.}
\label{fig:Lbol_vs_tAGN_cs} 
\end{figure}

In Figure \ref{fig:Lbol_vs_tAGN_cs}, we show the dependence of AGN bolometric luminosity 
on the duration of the AGN episode. The distribution peaks at $t_{\mathrm{AGN}} \approx 10^{7.5}$ yr and $L_{\mathrm{bol}} \approx 10^{42} \mathrm{ergs^{-1}}$, with objects with luminosities $L_{\mathrm{bol}} < 10^{44} \mathrm{ergs^{-1}}$ having a wide range of durations of the AGN episodes. However, the brightest objects at $L_{\mathrm{bol}} \approx 10^{48} \mathrm{ergs^{-1}}$ all have durations of $t_{\mathrm{AGN}} \approx 10^{6}$ yr with an anti-correlation between duration of the AGN episode and the AGN luminosity. This anti-correlation arises because in general, shorter AGN epsiodes lead to higher AGN luminosities.


\section{Evolution of the AGN luminosity function at z < 6}
\label{sec:comparison}

We first discuss the evolution of the predicted AGN luminosity function, as it is the simplest to predict, and then the AGN luminosity functions at different wavelengths, which depend on bolometric and obscuration corrections.

\subsection{Bolometric luminosity function}
\label{sec:bol_lf_func}

We present the predicted bolometric luminosity function compared to our observationally estimated bolometric luminosity function
constructed from multiwavelength data.
This observationally estimated bolometric luminosity function is described in Appendix \ref{app:visible_fraction}, and  
is compared to other observational estimates in Appendix \ref{app:visible_fraction}.

The model for SMBH evolution and AGN luminosity also involves some free parameters additional to those
in the galaxy formation model, as shown in Table \ref{tab:free_params}. We
have calibrated the values of $f_{\mathrm{q}}$ and $\mathrm{\eta_{Edd}}$, and found that the best-fitting values are those
adopted in \cite{fani12}. We show the effect of varying these parameters in Figures \ref{fig:bol_LF_varying_ftq}
and \ref{fig:bol_LF_varying_nEdd}. We also slightly adjust the values of $\alpha_{\mathrm{ADAF}}$ and $\alpha_{\mathrm{TD}}$ from 0.087 to 0.1.
This is for simplicity and to keep the values in line with MHD simulations \citep[e.g.][]{penna13}. The value of 
$\delta_{\mathrm{ADAF}}$ has been updated from \cite{fani12} (c.f. Section \ref{sec:lbol})

\begin{figure*}
\centering
\includegraphics[width=.9\linewidth]{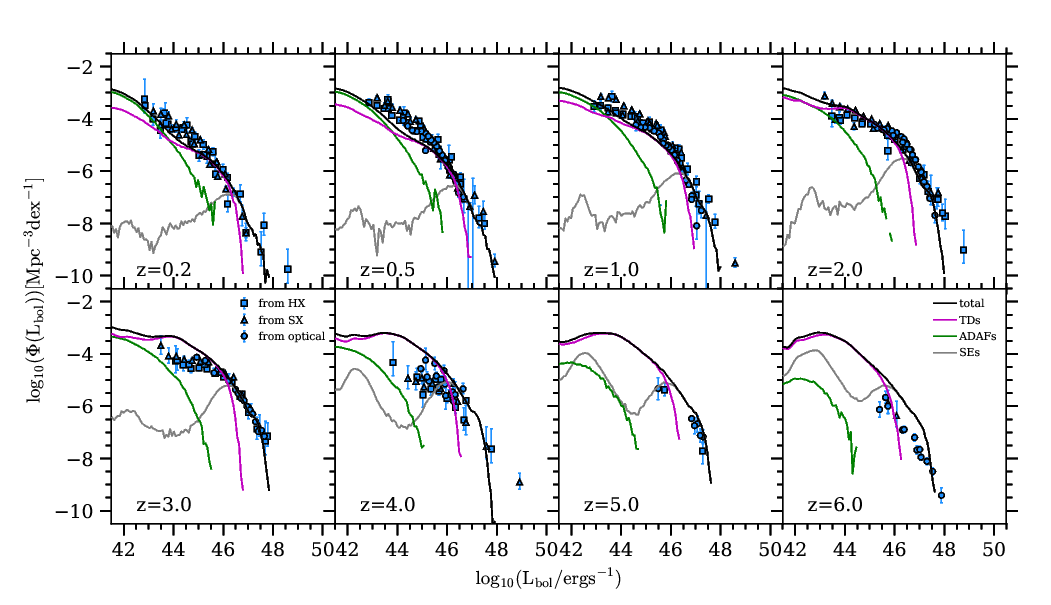}
\caption{The AGN bolometric luminosity function predicted by our model (black line, with grey shading showing 
the Poisson errorbars) compared to our bolometric luminosity function
constructed from the observations. We show the observational data indicating the wavelength of the data that was used to 
construct that particular point (squares - hard X-ray, triangles - soft X-ray, circles - optical). 
We split the total bolometric luminosity function by accretion mode into ADAFs (green), thin discs (purple) and super-Eddington objects (grey)}
\label{fig:L800_bol_lf_tdaf} 
\end{figure*}

In Figure \ref{fig:L800_bol_lf_tdaf}, the predictions (where the black line is the sum of the contributions from all accretion modes) compare well to the observational bolometric luminosity function across
the range of redshifts and for the luminosities shown. Exceptions include the faint end at high redshift 
where the model overpredicts the observations by 0.5 dex for $L_{\mathrm{bol}} < 10^{46} \mathrm{ergs^{-1}}$ for $z>4$,
and the faint end at low redshift where the model underpredicts the observations for $L_{\mathrm{bol}} < 10^{45} \mathrm{ergs^{-1}}$ and $z<0.5$ by 0.5 dex.
The underpredictions at the faint end at low redshift may be because the ADAF radiative accretion efficiency is lower than the thin disc
accretion efficiency, leading to lower luminosities (see Figure \ref{fig:L800_bol_etd} for a prediction using only a thin disc accretion efficiency for all
values of $\dot{m}$). Alternatively, this discrepancy might be resolved by
assuming an accretion timescale with a dependence on accreted gas mass or black hole mass. 
For a different model, \cite{shirakata18}
obtain a better fit to the hard X-ray luminosity function 
at low luminosity and low redshift by doing this.
In general, our model is a good match to these observations across a broad range.

We also show in Figure \ref{fig:L800_bol_lf_tdaf} the separate contributions to the AGN luminosity function from ADAFs ($\dot{m} < \dot{m}_{\mathrm{crit,ADAF}}$),
thin discs ($\dot{m}_{\mathrm{crit,ADAF}} < \dot{m} < \eta_{\mathrm{Edd}}$) and super-Eddington objects ($\dot{m} > \eta_{\mathrm{Edd}}$). 
At low redshift, ADAFs dominate the faint end ($L_{\mathrm{bol}} < 10^{44} \mathrm{ergs^{-1}}$), 
thin discs dominate at intermediate luminosities ($10^{44} \mathrm{ergs^{-1}} < L_{\mathrm{bol}} < 10^{46} \mathrm{ergs^{-1}}$) 
and super-Eddington objects dominate the bright end ($L_{\mathrm{bol}} > 10^{46} \mathrm{ergs^{-1}}$). As we go to higher redshift, the ADAFs 
contribution to the luminosity function decreases: for $0<z<2$ the evolution is not that strong, although the contribution from ADAFs 
at each luminosity decreases slightly as we increase $z$ in this range, whereas for $z>2$, the evolution in the ADAF population
is pronounced, and the number of ADAFs drops off sharply with increasing redshift. In contrast, the contribution from the thin disc population
increases until $z \approx 2$, after which it remains approximately constant. At $z<2$, there are not very many super-Eddington objects and so they make a fairly small contribution
to the luminosity function but their contribution increases at $z>2$. The distribution of super-Eddington objects
is bimodal, and for $z<4$, the higher luminosity peak has a higher number density, while for $z>4$, the lower luminosity peak has a higher number density.
The bimodality is not due to the bimodality in the fuelling modes, as all the super-Eddington objects are fuelled
by starbursts triggered by disc instabilities, but it seems to be caused by a bimodality in the bulge stellar mass.
We plan to explore this issue in more detail in future work.

\begin{figure*}
\centering
\includegraphics[width=.9\linewidth]{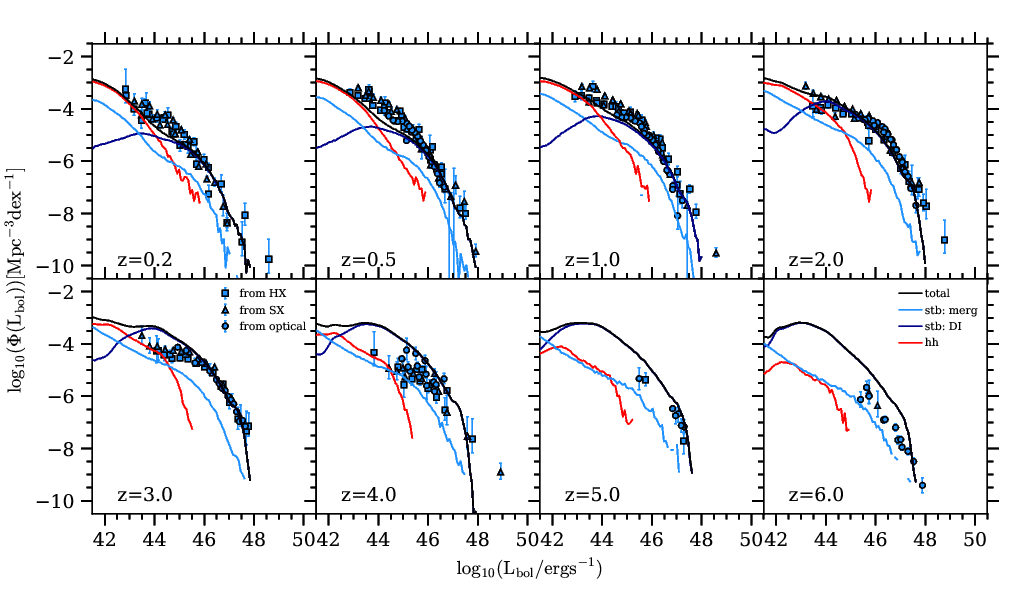}
\caption{The AGN bolometric luminosity function as Figure \ref{fig:L800_bol_lf_tdaf}, but split by the fuelling mode: starbursts triggered by mergers (light blue), starbursts triggered by
disc instabilities (dark blue), hot halo mode (red).}
\label{fig:L800_bol_lf_stbhh} 
\end{figure*}

In Figure \ref{fig:L800_bol_lf_stbhh} we split the AGN luminosity function by contributions from the hot halo mode, starbursts 
triggered by mergers and starbursts triggered by disc instabilities. At low redshift ($z<2$), the faint end is dominated by the hot halo mode, 
whereas the bright end is dominated by starbursts triggered by disc instabilities. Starbursts triggered by mergers
make a small contribution to the AGN bolometric luminosity function at low redshift. Starbursts triggered 
by disc instabilities typically have higher values of $\dot{M}$ and so higher luminosities 
compared to starbursts triggered by mergers, which is why they dominate the bright end.

The hot halo mode only operates in the most massive 
haloes, and so it only begins to significantly contribute to the AGN luminosity function for $z<3$.
The hot halo mode does not strongly evolve for $0<z<2$.
For $z>2$, starbursts triggered by disc instabilities dominate the AGN luminosity function, with starbursts
from mergers not significantly contributing. This implies that the inclusion of black hole growth via disc instabilities
is significant for reproducing AGN luminosity functions at high redshift.

A key aspect of the success of the \galform AGN model is the different channels of black hole growth, particularly 
the inclusion of disc instability triggered starbursts, that allow a good match to the AGN luminosity functions to be obtained. 
Other semi-analytic models do not necessarily
include disc instabilities, which may explain why they do not reproduce AGN properties particularly well at high redshift
\citep[e.g.][]{bonoli09,menci13,neistein14,enoki14}.
The effect of disc instabilities on the AGN predictions at $0<z<6$ is shown in Figure \ref{fig:L800_bol_stabledisk}
and the effect on galaxy properties is shown in \cite{lacey16}.

\begin{figure*}
\centering
\includegraphics[width=.8\linewidth]{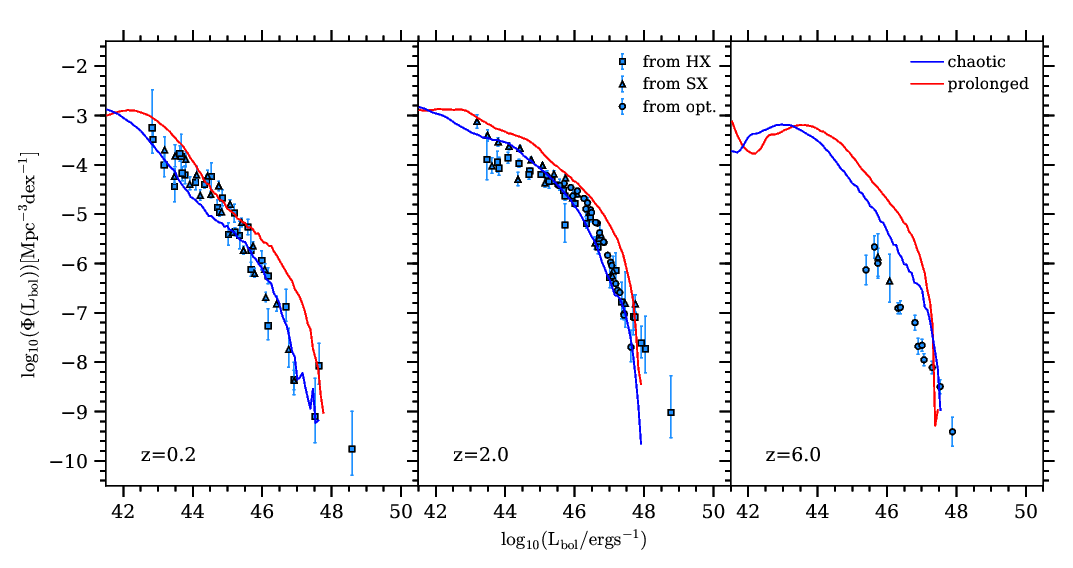}
\caption{The effect of changing between chaotic (blue) and prolonged (red) mode on the AGN bolometric luminosity function at $z=0.2,2,6$.}
\label{fig:bollf_chaotic_prolonged} 
\end{figure*}

We show the effect on the AGN bolometric luminosity function of changing between chaotic mode (our standard choice) and prolonged mode in Figure \ref{fig:bollf_chaotic_prolonged}. 
In the prolonged mode, SMBH spins are generally higher (see Figure \ref{fig:spin_vs_mbh}),  
which results in a higher radiative accretion efficiency leading to higher bolometric luminosities.\footnote{Note 
that the shape of the luminosity function changes little between the two models.}

\begin{figure*}
\centering
\includegraphics[width=.9\linewidth]{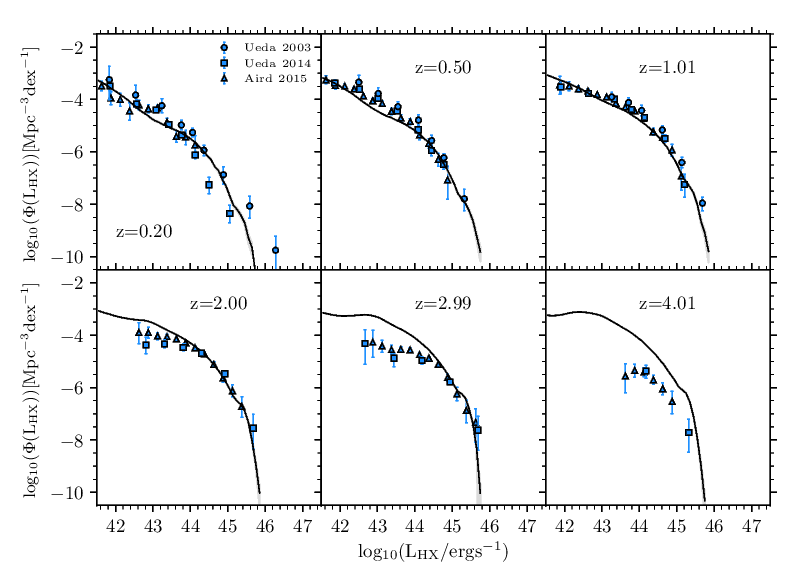}
\caption{The rest-frame hard X-ray luminosity function predicted by the model (black line) compared to observational studies
from \protect\cite{ueda03} (circles), \protect\cite{ueda14} (squares) and \protect\cite{aird15} (triangles).}
\label{fig:hx_LF} 
\end{figure*}

\begin{figure*}
\centering
\includegraphics[width=.9\linewidth]{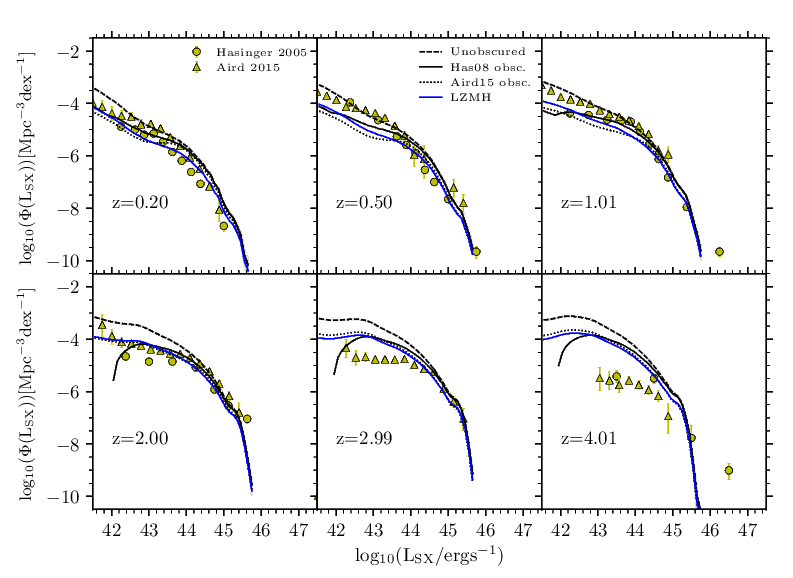}
\caption{The predicted rest-frame soft X-ray luminosity function compared to observations. The dashed black line shows the prediction
without accounting for absorption effects, the solid black line is the prediction using the \protect\cite{hasinger08} 
visible fraction,
the dotted black line is using the \protect\cite{aird15} visible fraction and the blue line is
using our observationally determined LZMH visible fraction.
The observations are \protect\cite{hasinger05} (circles) and \protect\cite{aird15} (triangles).}
\label{fig:sx_LF} 
\end{figure*}

\begin{figure*}
\centering
\includegraphics[width=.9\linewidth]{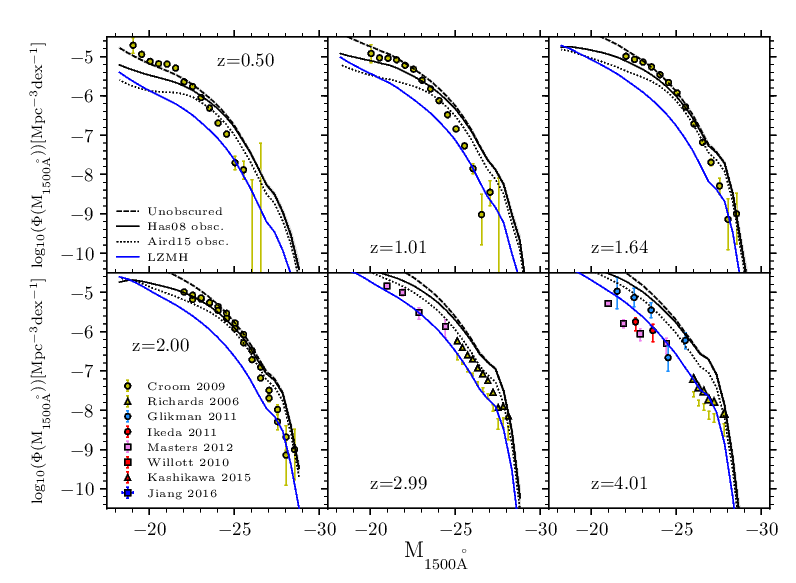}
\caption{The predicted rest-frame 1500$\angstrom$ luminosity function compared to observations which have been converted to 1500$\angstrom$. 
The dashed black line is the prediction 
without accounting for absorption effects, the solid black line is the prediction with the \protect\cite{hasinger08}
visible fraction, the dotted black line is
with the \protect\cite{aird15} visible fraction and the blue line is with my observationally determined LZMH visible fraction.
The observations are from SDSS DR3 \protect\cite{richards06} (yellow triangles), 2SLAQ+SDSS \protect\cite{croom09b} (yellow circles),
CFHQS+SDSS \protect\cite{willott10} (red squares), NDWFS+DLS \protect\cite{glikman11} (blue circles), the COSMOS field \protect\cite{ikeda11} (red circles), \protect\cite{masters12} (purple squares),
Subaru \protect\cite{kashikawa15} (red triangles) and SDSS Stripe 82 \protect\cite{jiang16} (blue squares).}
\label{fig:optical_LF} 
\end{figure*}

\begin{figure*}
\centering
\includegraphics[width=.8\linewidth]{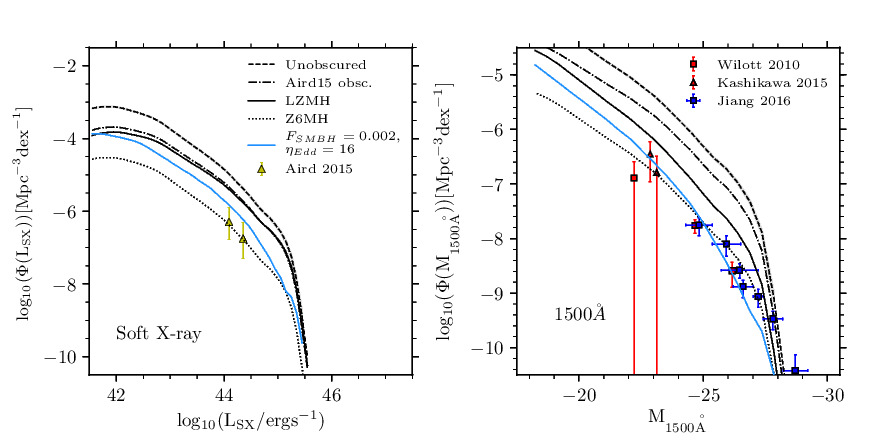}
\caption{The rest-frame soft X-ray luminosity function (left panel) and the rest-frame 1500$\angstrom$ luminosity function (right panel), both at $z=6$. We show predictions without obscuration (dashed black),
with the \protect\cite{aird15} visible fraction (dot-dash), with the `low z modified Hopkins' (LZMH) visible fraction with the standard model (black solid),
with the `$z=6$ modified Hopkins' (Z6MH) visible fraction (black dotted) and with the `low z modified Hopkins' visible fraction with the different parameters
(blue solid). The observations for the soft X-ray band are from \protect\cite{aird15} (yellow triangles), and for 1500$\angstrom$ are from \protect\cite{willott10} (red squares), \protect\cite{kashikawa15} (red triangles) and \protect\cite{jiang16} (blue squares).}
\label{fig:z6_sxlf_optical} 
\end{figure*}

\subsection{Luminosity functions at different wavelengths}

We use the SED template described in Section \ref{sec:bol_corr} and visible fractions described in Section \ref{sec:unobsc_frac}
to make predictions for the luminosity function in the rest-frame hard X-ray, soft X-ray and 1500$\angstrom$ bands. 
in Figure \ref{fig:hx_LF} we compare our hard X-ray predictions to observational data. The model is generally in good agreement with the 
observational data, particularly in the range $1 < z < 3$. For $L_{\mathrm{HX}} < 10^{44} \mathrm{ergs^{-1}}$ at $z<0.5$, the model underpredicts the 
observations by about 0.5 dex, and for $L_{\mathrm{HX}} < 10^{44} \mathrm{ergs^{-1}}$ at $z>3$, the model overpredicts the observations by about
1 dex. The former discrepancy corresponds to the model bolometric luminosity function underpredicting the observations 
in the same redshift and luminosity regime, and the latter also cooresponds to the bolometric luminosity function
slightly overpredicting the observational estimates in that regime, but may also be influenced by our assumption 
that there is no obscuration for hard X-ray sources. This assumption may be not valid
for the high redshift Universe; more observations are needed to constrain the obscuration effect on hard X-rays.

Our soft X-ray predictions are compared to observations in Figure \ref{fig:sx_LF}.
The predicted luminosity function without taking into account obscuration is shown alongside the model with the 
visible fractions of \cite{hopkinsrh07}, \cite{hasinger08}, \cite{aird15} and our observationally determined LZMH model. 
The luminosity functions with different visible fractions are very similar
except for $L_{\mathrm{SX}} < 10^{44} \mathrm{ergs^{-1}}$. The LZMH model fits best to the observations in the range $1 < z < 2$.
At higher redshifts and lower luminosities the visible fraction in the \cite{hasinger08} model drops to zero,
which causes the corresponding drop off in the luminosity function for that obscuration model.

Our 1500$\angstrom$ predictions are shown in Figure \ref{fig:optical_LF} compared to observational estimates.
These have been converted to 1500$\angstrom$ - 
the conversions made are detailed in Appendix \ref{app:optical_magnitudes}.
There is a strong dependence of the predictions on the assumed obscuration model. Our predictions are a good fit to observations at $z \approx 2$ if we adopt the \cite{hasinger08}
visible fraction, whereas our observationally determined LZMH model fits best for $z \approx 4$.
The reason for this difference is likely to be because \cite{hasinger08} fitted their obscuration
model at lower redshift whereas we are trying to fit for $z=0-6$ with our LZMH visible fraction.
Therefore, unsurprisingly, the different visible fractions are likely to fit better in different redshift ranges.

We present the soft X-ray and optical luminosity functions at $z=6$ in Figure \ref{fig:z6_sxlf_optical}. The predicted soft X-ray luminosity function exceeds the observations at $z=6$ as a result of the model bolometric luminosity function overpredicting the observations. For the optical luminosity function, while the model gives an acceptable fit to observations of the optical luminosity function at $z=4$, it overpredicts the number of 
AGN compared to the observed luminosity function at $z=6$. This is a result of the model not strongly evolving in the redshift 
interval $z=4-6$, while the observations
indicate a stronger evolution in this redshift interval \citep{jiang16}. These discrepancies could be due to a variety of reasons.
We suggest two possible explanations for this discrepancy and two corresponding variants on the model which provide a better 
fit to the observations at $z=6$.

Firstly, the discrepancy could be due to the obscuration model. At $z=6$ the visible fraction is not constrained by any observations,
and so in Figure \ref{fig:z6_sxlf_optical} we present predictions with a lower visible fraction at $z=6$, which give a better fit to the $z=6$ optical luminosity 
function. We show predictions for the standard model with two obscuration models:
the LZMH visible fraction and the Z6MH visible fraction (c.f. Section \ref{sec:unobsc_frac}).
The Z6MH visible fraction needed to fit $z=6$ is about a quarter of the LZMH visible fraction at $z<6$. Thus $z>6$ QSOs could be 
much more obscured than $z<6$ QSOs. 

Secondly, the discrepancy could be due to black hole accretion being less efficient at high redshift. While the model for black hole accretion has been calibrated at low redshift, the conditions for black hole accretion could be different at higher redshift.
We therefore present a model with parameters that have been modified compared to the original calibration on observed
data at low redshift. We change the parameter $f_{\mathrm{BH}}$, which sets the fraction of mass accreted onto a black hole
in a starburst event and the parameter $\eta_{\mathrm{Edd}}$, which controls the degree of super-Eddington
luminosity suppression. In the fiducial model, $f_{\mathrm{BH}}=0.005$ and $\eta_{\mathrm{Edd}}=4$. $f_{\mathrm{BH}}=0.002$ and $\eta_{\mathrm{Edd}}=16$ give a better fit to the observations 
of the $1500\angstrom$ luminosity function at $z=6$ in 
Figure \ref{fig:z6_sxlf_optical}. However, we note that $\eta_{Edd}=16$ means that there is very little super-Eddington luminosity suppression,
whereas the `slim disc' model for super-Eddington sources predicts significant super-Eddington luminosity suppression. We refer to this model as the
`low accretion efficiency model'. In this model we use the LZMH visible fraction.

Both of these alternative models are in better agreement with observations of the $1500 \angstrom$ AGN luminosity function at $z=6$ than our standard model, and so we will use them for future studies inestigating AGN observed in future surveys.

\subsection{Comparison with hydrodynamical simulations}

An alternative theoretical approach for simulating galaxy formation is hydrodynamical simulations. A few of these simulations have been used to make predictions for the evolution of AGN luminosity functions through time. We give a brief comparison to some of these here.

The bolometric luminosity function predicted by the model in \cite{hirschmann14} over the redshift range $0 < z< 5$ is shown in their Figure 8. When compared to \cite{hopkinsrh07}, their model is a good fit to the observations at $z=0.1$, but overpredicts the observations at the faint end at $z=2$, and underpredicts the observations at $z=5$. When comparing their results to the model presented in this paper (c.f. Figure \ref{fig:L800_bol_lf_tdaf}), our model agrees similarly well with the observations for $z<2$, and with better agreement to the observations for $z>2$. For example, at $z=4$, at $\lbol = 10^{46} \mathrm{ergs^{-1}}$ (around the knee of the luminosity function at this redshift), our model agrees within 0.5 dex with the observed bolometric luminosity function, whereas the model of \cite{hirschmann14} underpredicts the observed bolometric luminosity function by 1 dex at this redshift and luminosity.
The hard X-ray luminosity function predicted by EAGLE in  \cite{rosasguevara16} is compared to the observational estimate of \cite{aird15} over the redshift range $0<z<5$ in their Figure 7. Their model fits well to the observations at $z=0$, but by $z=1$, the slope of the luminosity function in their work is steeper than the observations. The model in this paper is in similar agreement for $z<1$, and in better agreement with the observations for $z>1$. For example, at $z=2$, at $\log(L_{\mathrm{HX}})= 10^{44} \mathrm{ergs^{-1}}$ (around the knee of the luminosity function at this redshift), our model agrees within 0.5 dex with the observations, whereas the model of \cite{rosasguevara16} underpredicts the observations by about 1 dex.
Finally, \cite{weinberger18} compare the bolometric luminosity function from IllustrisTNG to \cite{hopkinsrh07} in the redshift range $0<z<5$. Their model underpredicts the observations at the faint and bright end of the bolometric luminosity function and overpredicts the observations at intermediate luminosities at $z=0.5$, and overpredicts the observations at all luminosities at $z=3$. Around the knee of the luminosity function at $z=3$ ($\lbol = 3 \times 10^{46} \mathrm{ergs^{-1}}$), our model agrees within 0.5 dex with the observations, whereas the model of \cite{weinberger18} overpredicts the observations by 0.5 dex.

Overall, the AGN luminosity functions from the hydrodynamical simulations do not agree as well to the observational estimates as this model. The reasons for the differences in the AGN luminosity functions may be because the black hole mass accretion rates are calculated differently - in these simulations the Bondi-Hoyle approximation is used, as opposed to the calculation in Section \ref{sec:lbol} used in this work.



\section{Conclusions}
\label{sec:conclusions}


Understanding the evolution of AGN across cosmic time has been of interest ever since they were discovered. AGN have also been shown to be important
in how galaxies evolve through AGN feedback. 
However, many uncertainties remain,
such as the nature of the physical processes involved in AGN feedback.

We present predictions for the evolution of SMBHs and AGN at $0<z<6$ using a high volume, high
resolution dark matter simulation (P-Millennium) populated with galaxies using the semi-analytic model of galaxy formation \galform.
This updated scheme for the SMBH spin evolution is used within the \cite{lacey16} \galform model as updated by \cite{baugh18} for the P-Millennium simulation.
The \cite{lacey16} model has been shown to reproduce a large number of observable galaxy properties over an unprecedented wavelength 
and redshift range.
The model that we use incorporates an updated prescription for SMBH spin evolution:
for these predictions we have assumed SMBH spin evolving in a `chaotic
accretion' scenario in which the angle between the accretion disc and the SMBH spin randomises 
once a self-gravity mass of gas has been consumed.

We then calculated AGN bolometric luminosities from the SMBH mass accretion rate, taking into 
account the SMBH spin and the different radiative efficiencies for
different accretion regimes (ADAFs, thin discs, super Eddington objects). Then using 
a template SED and different obscuration models we derived AGN luminosities in the hard X-ray, soft X-ray and optical/UV (1500$\angstrom$) 
bands.

The model predictions are consistent with both the observed black hole mass functions and SMBH mass versus bulge 
mass correlations. We present the spin distribution of SMBHs in the simulation, for the chaotic and prolonged modes of accretion, and their evolution for $0<z<6$. The median SMBH spin in both the chaotic and prolonged modes evolves very little. For the prolonged mode, the scatter in the SMBH spin distribution decreases with increasing redshift. We also present examples of the evolution of spin and mass for typical SMBHs, and find that for most masses the evolution is similar, except at the highest masses, $\mbh > 10^8 M_{\odot}$, where mergers cause the SMBHs to be spun up to higher spin values.

We compare the AGN luminosity functions in the redshift range $0 < z < 6$
to a wide range of observations at different wavelengths. 
The model is in good agreement with the observations. We split the luminosity functions by accretion mode 
(ADAFs, thin discs, super-Eddington objects) and by fuelling mode (hot halo or starbursts triggered by disk instabilities or 
mergers) to see the relative contributions. At low redshifts, $z<2$, and low luminosities, $\lbol < 10^{43} \mathrm{ergs^{-1}}$, the ADAF contribution dominates 
but at higher luminosities and higher redshifts, the thin disc and super-Eddington objects dominate the 
luminosity function. Hot halo mode fuelled accretion dominates at $z<3$, and $\lbol < 10^{44} \mathrm{ergs^{-1}}$, but at higher redshift 
and higher luminosity, starbursts triggered by disc instabilities dominate the luminosity function.

There are many natural continuations from this work. 
We have already mentioned that we always assume a quasar SED for our bolometric corrections,
while in reality we have a variety of AGNs having different accretion rates in different accretion regimes, which will 
have different SED shapes \cite[e.g.][]{jin12c}. Using different template SEDs
for different regimes may allow the model to predict luminosity functions in better agreement 
with the observations.
Secondly, we could more thoroughly explore the dependence of the model on the SMBH spin evolution model 
used e.g. investigating the dependence of the results on the size of the increments in mass used in the SMBH accretion calculation.
Finally, in this paper we do not show radio luminosity functions
- given that AGN jets are observed to have a strong effect on their host galaxies and given that these jets 
emit at radio wavelengths via synchrotron emission, an investigation into radio emission would also be important
for understanding the role of AGN in galaxy evolution.

\section*{Acknowledgements}

This work was supported by the Science and Technology facilities Council ST/L00075X/1.
AJG acknowledges an STFC studentship funded by STFC grant ST/N50404X/1. 
CL is funded by an Australian Research Council Discovery Early Career Researcher Award 
(DE150100618) and by the Australian Research Council Centre of Excellence for All Sky 
Astrophysics in 3 Dimensions (ASTRO 3D), through project number CE170100013.
This work used the DiRAC Data Centric system at Durham University, operated by
the Institute for Computational Cosmology on behalf of the STFC DiRAC HPC
Facility (www.dirac.ac.uk). This equipment was funded by BIS National
E-infrastructure capital grant ST/K00042X/1, STFC capital grants ST/H008519/1
and ST/K00087X/1, STFC DiRAC Operations grant ST/K003267/1 and Durham
University. DiRAC is part of the National E-Infrastructure.




\bibliographystyle{mnras}
\bibliography{references}




\appendix

\section{Effects of varying SMBH seed mass}
\label{app:seed_mass}

\begin{figure}
\centering
\includegraphics[width=\linewidth]{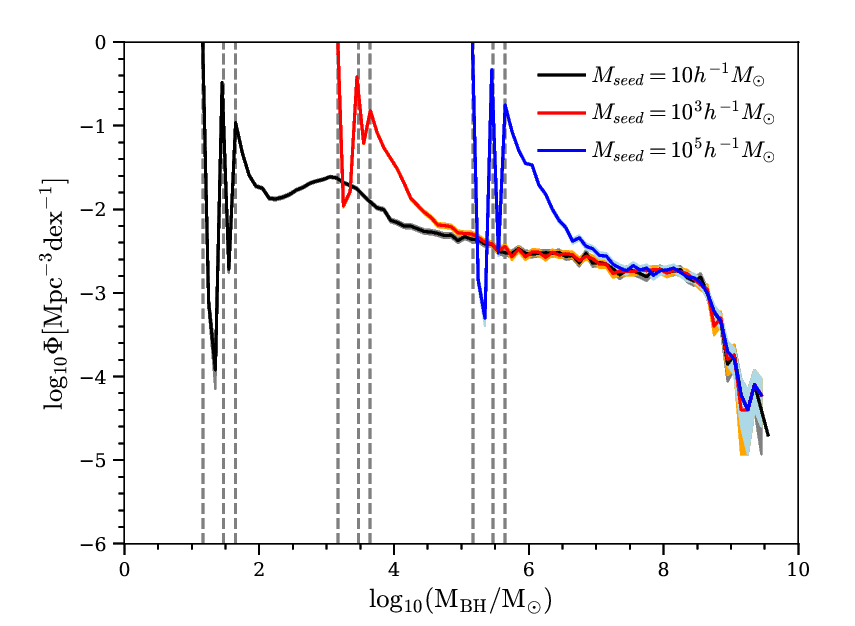}
\caption{The black hole mass function at $z=0$ for seed masses of $10h^{-1} M_{\odot}$ (black),
$10^{3}h^{-1} M_{\odot}$ (red) and $10^{5}h^{-1} M_{\odot}$ (blue).}
\label{fig:Seed_BHMF_z0} 
\end{figure}

\begin{figure}
\centering
\includegraphics[width=\linewidth]{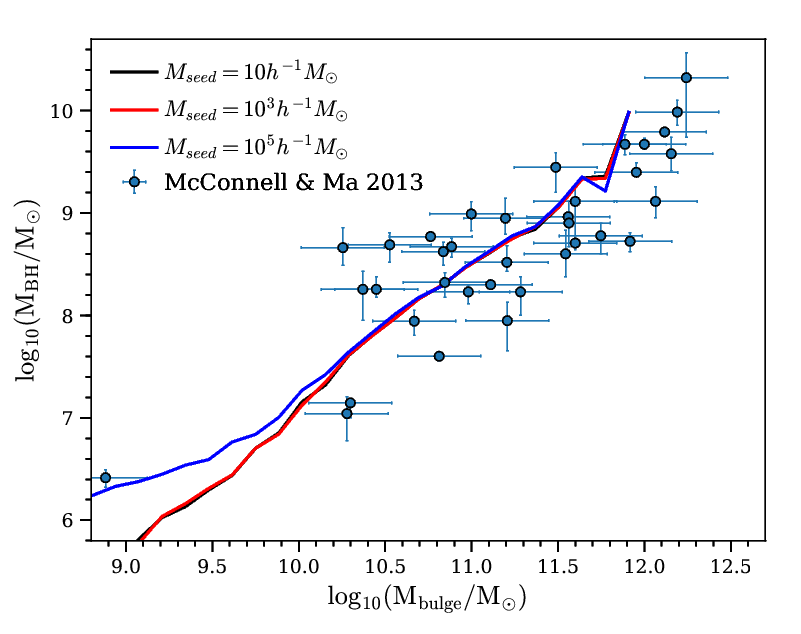}
\caption{The predicted SMBH mass versus SMBH mass relation at $z=0$ for seed masses of $10h^{-1} M_{\odot}$ (black),
$10^{3}h^{-1} M_{\odot}$ (red) and $10^{5}h^{-1} M_{\odot}$ (blue) compared to \protect\cite{mcconnellma13}.}
\label{fig:Seed_MBH_Mbulge} 
\end{figure}

In Figure \ref{fig:Seed_BHMF_z0} we show the effect of varying the SMBH seed mass on the black hole 
mass function at $z=0$. 
We show plots for SMBH seed masses of $10 h^{-1} M_{\odot}$ (the default value), $10^{3} h^{-1} M_{\odot}$ and $10^{5} h^{-1} M_{\odot}$.
Generally the black hole mass function reaches a converged value at about 100 times the black hole
seed mass.
We also plot as vertical lines $\mbh = M_{\mathrm{seed}}$, $\mbh = 2 \times M_{\mathrm{seed}}$ and $\mbh = 3 \times M_{\mathrm{seed}}$.
It can be seen that the spikes in the black hole mass function occur at these values due to SMBH seeds 
merging with other SMBH seeds.

This convergence in properties at around 100 times the seed mass can also be seen in Figure \ref{fig:Seed_MBH_Mbulge}, where the median of the SMBH mass versus bulge mass relation 
for seeds of mass $10^{5} h^{-1} M_{\odot}$
only converges with that for the other seed masses for SMBH masses above around $10^{7} M_{\odot}$.

\section{Calculating broad-band optical magnitudes for AGN}
\label{app:optical_magnitudes}

We define the filter-averaged luminosity per unit frequency for a filter R in the observer frame at redshift $z$ as:

\begin{equation}
 < L_{\nu} >_{R}^{(z)} = \frac{\int L_{\nu}((1+z)\nu_{o}) R(\nu_{o}) d \nu_{o}}{\int R(\nu_{o}) d \nu_{o}}, \label{eq:l_nu_r_z}
\end{equation}

\noindent
where $L_{\nu}(\nu)$ is the luminosity per unit frequency in the rest frame, $R(\nu_{o})$ is the response function of the filter at observed frequency $\nu_{o}$. 
The absolute magnitude in the AB system in the observer frame band defined by the filter R for redshift $z$, is then defined as:

\begin{equation}
 M_{AB, R}^{(z)} = -2.5 \mathrm{log}_{10} \Big( \frac{< L_{\nu} >_{R}^{(z)}}{L_{\nu_{o}}} \Big), \label{eq:abs_mag}
\end{equation}

\noindent
where $L_{\nu_{o}} = 4 \pi (10 \mathrm{pc}^2)\times f_{\nu_{o}}$ with $f_{\nu_{o}} = 3631$Jy, the flux corresponding to
an apparent AB magnitude of 0, and $L_{\nu_{o}}$ the corresponding luminosity per unit frequency for an absolute AB magnitude of 0. 
We remind readers that monochromatic AB (Absolute Bolometric) apparent magnitudes are defined using the following relation \citep{okegunn83}:

\begin{equation}
 m_{\mathrm{AB}}(\nu) = -2.5 \mathrm{log}_{10} \Big( \frac{f_{\nu}}{f_{\nu_{o}}} \Big),
\end{equation}

\noindent
where $f_{\nu}$ is the observed flux of the source, which is related to the luminosity per unit frequency in the rest-frame
of the object as:

\begin{equation}
 f_{\nu}(\nu_{o}) = \frac{(1+z)L_{\nu}((1+z)\nu_{o})}{4 \pi d_{L}^2}.
\end{equation}

\noindent The apparent and observer frame absolute magnitudes for a filter R are then related by

\begin{equation}
\begin{split}
 m_{AB}(\nu) = -2.5 \mathrm{log}_{10} \Big( \frac{< L_{\nu} >_{R}^{(z)}}{L_{\nu_{o}}} \Big) - 2.5\mathrm{log}_{10}(1+z) \\ + 5 \mathrm{log}_{10} (d_{\mathrm{L}} / 10 \mathrm{pc}). \label{eq:m_app_abs}
 \end{split}
\end{equation}

We then use the following formulae to convert the observational data from the different wavelengths given to rest-frame 
wavelength 1500$\angstrom$. 
Note that we are are only comparing continuum luminosities in this study, which is consistent with the 
\cite{marconi04} template used throughout this paper. The data presented in the studies that we use have the contribution from the emission lines 
removed and so this is an appropriate comparison.
The results presented in \cite{richards06} are given in the K-corrected SDSS i band at $z=2$, which we write as $M'_{i}(z=2)$. This is given by 
$M'_{i}(z=2) = M_{i}(z=2) - 2.5\mathrm{log}(1+z)$,
where we define $M_{i}(z=2)$ as the absolute magnitude at the rest-frame wavelength corresponding to the observed i-band at $z=2$,
as in equations (\ref{eq:abs_mag}) and (\ref{eq:m_app_abs}).
To convert from $M_{i}(z=2)$ to 1500$\angstrom$, we follow \cite{richards06} by using $L_{\nu} \propto \nu^{\alpha_{\nu}}$ but using a spectral index value 
of $\alpha_{\nu} = -0.44$ from \cite{marconi04} instead of $\alpha_{\nu} = -0.5$ in \cite{richards06}. 
First we convert from $M'_{i}(z=2)$ to $M_{i}(z=0)$ using equations (\ref{eq:l_nu_r_z}) and (\ref{eq:abs_mag}):

\begin{align}
\begin{split}
 M_{i}(z=0) &= M'_{i}(z=2) + 2.5(1+\alpha_{\nu}) \mathrm{log}(1+2) \\
 &= M'_{i}(z=2) + 0.668, \label{eq:m_z0_z2}
\end{split} 
\end{align}

\noindent 
where $M_{i}(z=0)$ is the absolute magnitude at the central wavelength of the rest-frame i-band ($7471\angstrom$)
corresponding to equation (\ref{eq:abs_mag}) for $z=0$. 
Then we relate $M_{i}(z=0)$ to the absolute magnitude at rest-frame $1500\angstrom$, $M_{1500}$, to give the conversion to $M'_{i}(z=2)$:

\begin{align}
\begin{split}
  M_{1500} &= M_{i}(z=0) + 2.5 \alpha_{\nu} \mathrm{log}_{10} \Big( \frac{1500 \angstrom}{7471 \angstrom} \Big), \\
  &= M_{i}(z=0) + 0.767, \\
  &= M'_{i}(z=2) + 1.435.
\end{split}  
\end{align}

\noindent 
where in the last line we used equation (\ref{eq:m_z0_z2}).
\cite{jiang09, willott10, ikeda11, masters12, kashikawa15} report observed absolute continuum magnitudes, $M_{1450}$, corresponding to rest frame $1450 \angstrom$. These absolute magnitudes are defined without the extra redshift factor
included in the \cite{richards06} definition. These absolute magnitudes at $1450 \angstrom$, $M_{1450}$, can be converted to 1500$\angstrom$ using:

\begin{align}
\begin{split}
 M_{1500} &= M_{1450} + 2.5 \alpha_{\nu} \mathrm{log}_{10} \Big( \frac{1500 \angstrom}{1450 \angstrom} \Big), \\
 &= M_{1450} - 0.016. \label{eq:m1500_m1450}
\end{split}
\end{align}

\noindent
Finally \cite{croom09b} report observations in the SDSS g-band ($4670 \angstrom$) K-corrected to $z=2$, so we use the correction in their paper:

\begin{equation}
 M'_{g}(z=2) = M'_{i}(z=2) + 2.5 \alpha_{\nu} \mathrm{log} \Big( \frac{4670 \angstrom}{7471 \angstrom} \Big),
\end{equation}

\noindent
and combine it with the above relation to give:

\begin{equation}
 M_{1500} = M'_g(z=2) + 1.211.
\end{equation}

\section{Visible and obscured fractions for AGN}
\label{app:visible_fraction}

\begin{figure*}
\centering
\includegraphics[width=.7\linewidth]{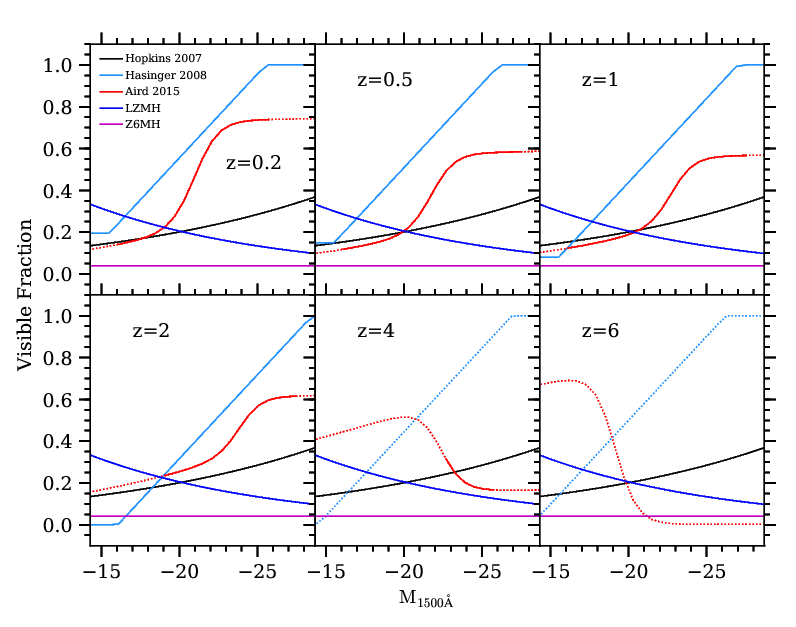}
\caption{Comparing the visible fractions for rest-frame $1500 \angstrom$ for different obscuration models. Shown are \protect\cite{hopkinsrh07} (black),
\protect\cite{hasinger08} (light blue), \protect\cite{aird15} (red), the LZMH model (dark blue) and the Z6MH model (purple).
The solid lines for the observational visible fractions indicate the ranges where there is observational data,
while the dotted lines indicate ranges where a functional form has been extrapolated.}
\label{fig:Obscuration_fraction_comp_opt} 
\end{figure*}

\begin{figure*}
\centering
\includegraphics[width=.7\linewidth]{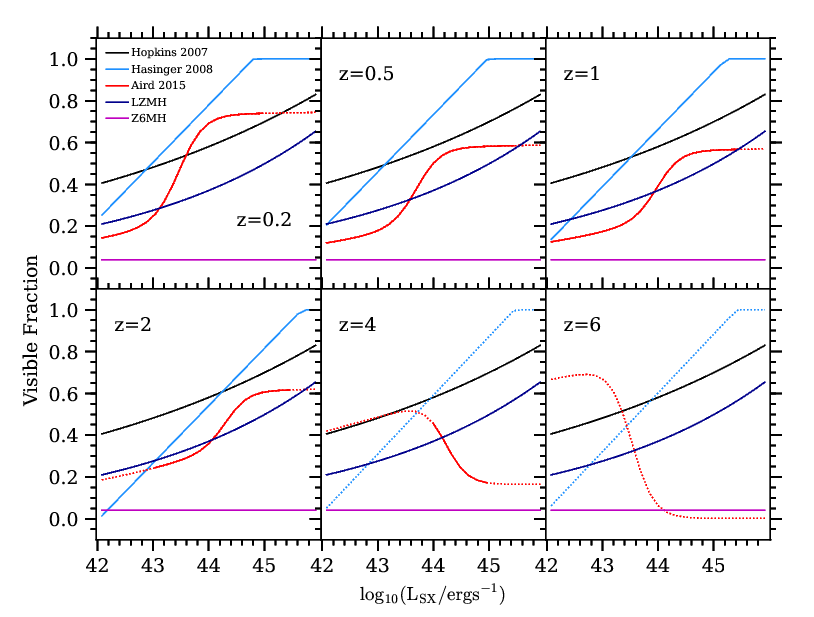}
\caption{The same as the previous plot, but for rest-frame soft X-rays.}
\label{fig:Obscuration_fraction_comp_SX} 
\end{figure*}

\begin{figure*}
\centering
\includegraphics[width=.7\linewidth]{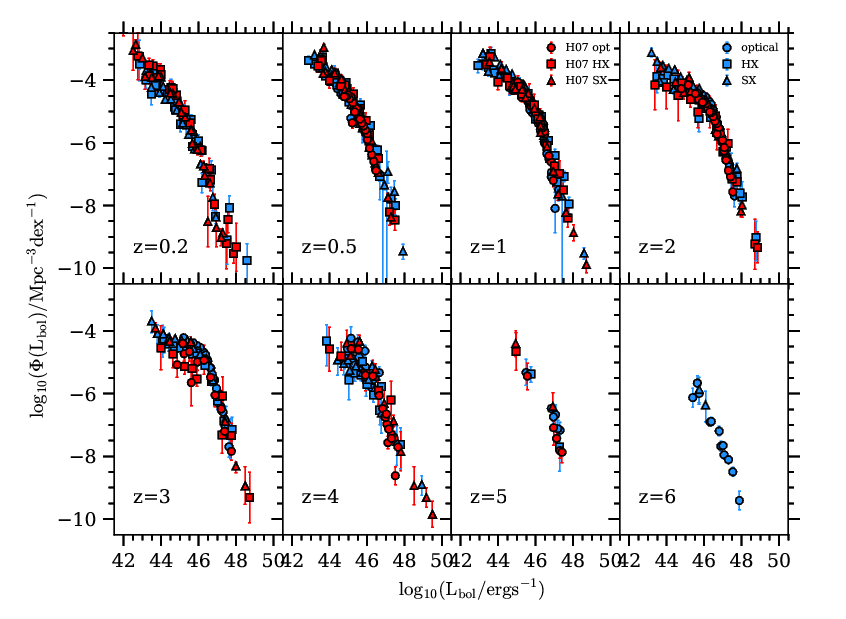}
\caption{The bolometric luminosity function derived in this work (blue) by using the \protect\cite{marconi04} bolometric corrections,
and by varying the coefficients of the visible fractions to obtain a bolometric luminosity function
with the smallest scatter between points derived from data at different wavelengths, compared to the \protect\cite{hopkinsrh07} bolometric luminosity function (red).
The \protect\cite{hopkinsrh07} bolometric luminosities have been multiplied by 7.9/11.8 to account for the different 
SED template used (see text).}
\label{fig:My_Bol_LF_with_Hop07} 
\end{figure*}

\begin{figure*}
\centering
\includegraphics[width=.7\linewidth]{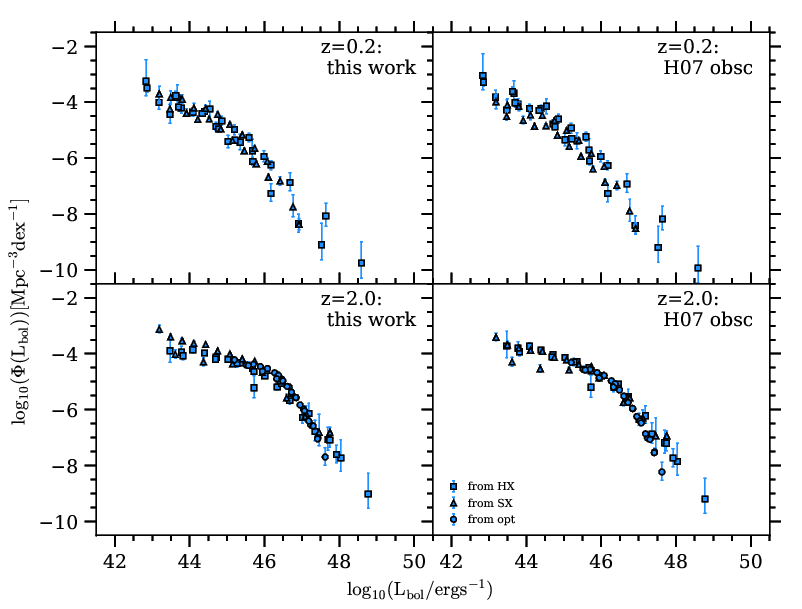}
\caption{Comparing the effect of using different obscuration models on the constructed bolometric luminosity functions. The left panels are obtained using the obscuration model presented in this work, while the right panels use
the obscuration model of \protect\cite{hopkinsrh07}. The upper panels are for $z=0.2$ and the lower panels are for $z=2$.}
\label{fig:Compare_obscuration_on_bol} 
\end{figure*}

The AGN visible fractions (the fraction of sources at a particular luminosity and redshift that are unobscured) 
derived in this paper have been estimated by constructing an observational bolometric luminosity function from 
observed luminosity functions at X-ray and optical wavelengths.
These luminosities were converted to bolometric using the \cite{marconi04} AGN SED, and then the observed number densities were 
converted to total number densities using visible fractions
of a functional form similar to \cite{hopkinsrh07} dependent only on $\lbol$ (c.f. equation (\ref{eq:hop_vis_frac})). We assumed that
there is no obscuration for hard X-ray wavelengths. The coefficients in the expressions for the visible fractions
were then selected (c.f. equations (\ref{eq:LZMH_opt}), (\ref{eq:LZMH_SX}) and (\ref{eq:Z6MH})) so as to minimise the scatter in the estimated bolometric luminosity function.

To construct a bolometric luminosity function from multiple sets of observations in different wavebands,
different authors use different template SEDs.
Some authors include reprocessed radiation
from dust (its inclusion causes an `IR bump' in the SED) whereas some do not. 
Including reprocessed radiation gives observed bolometric luminosities, whereas not including the IR
bump gives intrinsic bolometric luminosities. The intrinsic bolometric luminosities are isotropic, 
while the observed bolometric luminosities are not isotropic because the obscuring torus is not isotropic. 
The observed bolometric luminosity functions of \cite{hopkinsrh07} are given in terms of observed rather than intrinsic bolometric luminosities, so when we compare with these, we multiply the luminosities of \cite{hopkinsrh07} by a factor $7.9/11.8$ \citep[c.f.][]{marconi04} to account for this effect.  

We show a comparison of the different obscuration models at $1500\angstrom$ in Figure 
\ref{fig:Obscuration_fraction_comp_opt} and at soft X-ray energies in Figure \ref{fig:Obscuration_fraction_comp_SX}.
The values from different studies are not all on a single curve, and so there is clearly still some uncertainty in the visible fraction.

Our bolometric luminosity function is shown compared to the bolometric luminosity functions 
estimated in \cite{hopkinsrh07} in Figure \ref{fig:My_Bol_LF_with_Hop07}, and the two are 
in agreement. The bolometric luminosity 
function derived in this work is also similar to that determined by \cite{shankar09}. 

Our observationally estimated visible fractions are redshift independent by construction. We have explored whether a 
better fit could be obtained by including a redshift dependence. To obtain a better fit, the visible fraction needs to
increase and then decrease with redshift \cite[c.f. the redshift dependence derived by][]{aird15},
but even with a functional form to allow this, the scatter in the bolometric luminosity function
was only slightly less than for redshift independent versions of the visible fraction.

To quantify the effect of using the new visible fraction derived in this paper, we compare the bolometric luminosity 
function derived using the \cite{hopkinsrh07} visible fraction, to the bolometric luminosity function
derived using the visible fraction presented in this paper, in Figure \ref{fig:Compare_obscuration_on_bol}.
The new visible fraction does improve the constructed bolometric luminosity function, this reduction in scatter 
can be seen particularly at $L_{\mathrm{bol}} \sim 10^{44} \mathrm{ergs^{-1}}$ at $z=0.2$ and 
at $L_{\mathrm{bol}} \sim 10^{48} \mathrm{ergs^{-1}}$ at $z=2$.

\section{The effect of the time averaging method}
\label{app:time_averaging}

\begin{figure*}
\centering
\includegraphics[width=.7\linewidth]{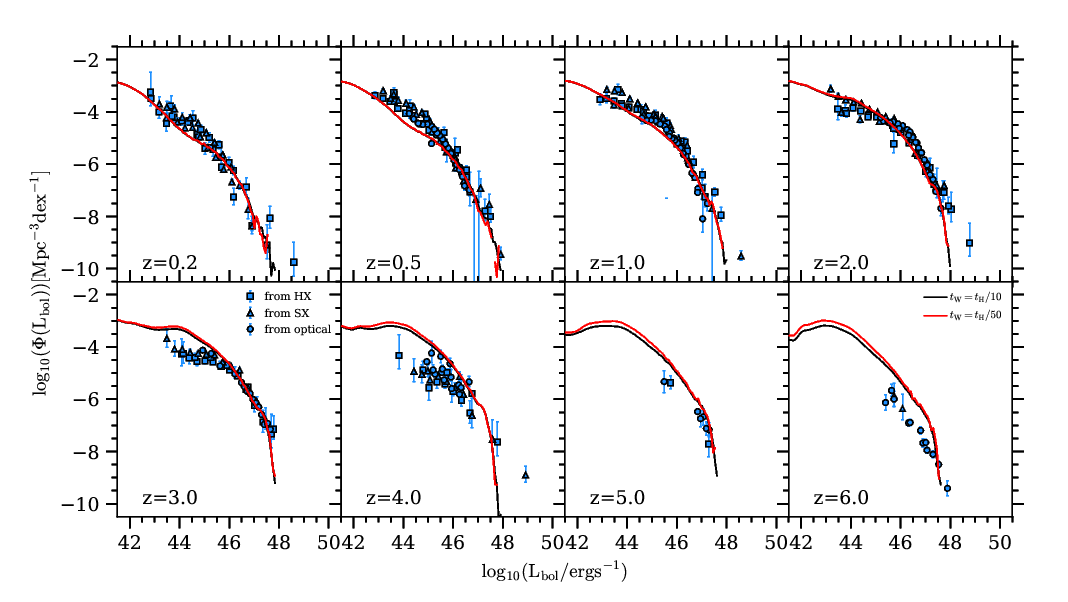}
\caption{Exploring the effect on the AGN bolometric luminosity function of varying $\Delta t_{\mathrm{window}}$, shown 
are $\Delta t_{\mathrm{window}}=t_{\mathrm{H}}/10$ (black) and $\Delta t_{\mathrm{window}}=t_{\mathrm{H}}/50$ (red).}
\label{fig:t_window_10_50} 
\end{figure*}

\begin{figure*}
\centering
\includegraphics[width=.7\linewidth]{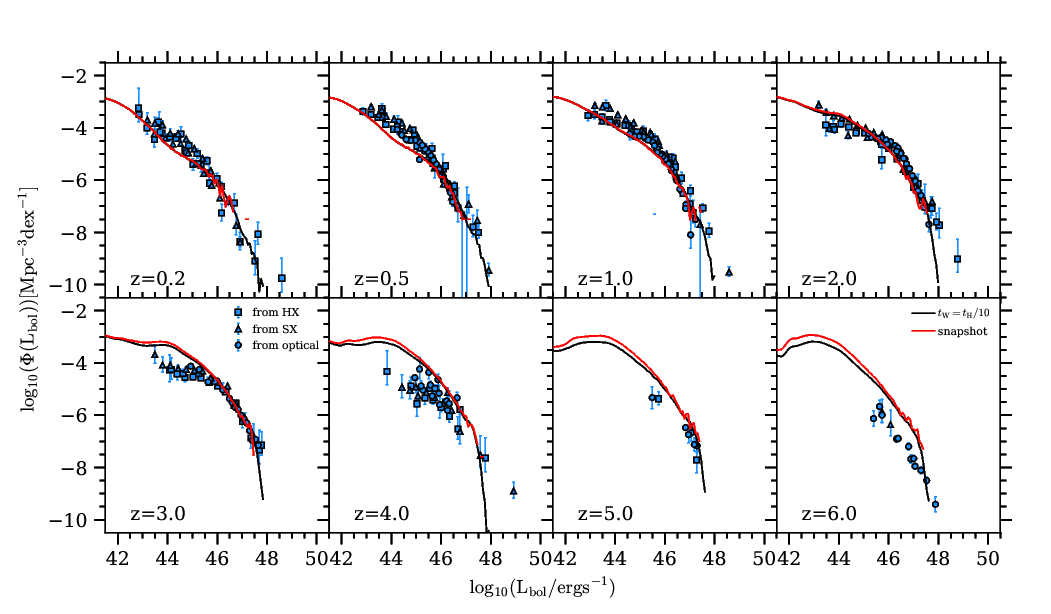}
\caption{Exploring the effect on the AGN bolometric luminosity function of varying $\Delta t_{\mathrm{window}}$, shown 
are $\Delta t_{\mathrm{window}}=t_{\mathrm{H}}/10$ (black) and using the snapshot luminosities (red).}
\label{fig:t_window_10_snapshot} 
\end{figure*}

In this appendix, we show the effect of varying $\Delta t_{\mathrm{window}}$ on the AGN luminosity function, as introduced in Section 
\ref{sec:time_av}, and compare the luminosity function obtained using the time averaging method in 
Section \ref{sec:time_av} to a luminosity function constructed using the snapshot luminosities.
In Figure \ref{fig:t_window_10_50}, the predicted luminosity function with a value of $\Delta t_{\mathrm{window}}=t_{\mathrm{H}}/10$
(the standard model), is compared to the predicted luminosity function with a value of $\Delta t_{\mathrm{window}}=t_{\mathrm{H}}/50$. The two are very similar, except at low luminosities at high redshift, where there is a slight difference. The similarity shows that the value of $\Delta t_{\mathrm{window}}$ adopted does not strongly affect the luminosity function.
In Figure \ref{fig:t_window_10_snapshot}, the predicted luminosity function 
with a value of $\Delta t_{\mathrm{window}}=t_{\mathrm{H}}/10$ is compared to the luminosity function where only the snapshot 
luminosities are used to construct the luminosity function. It can be seen how the time averaging method allows predictions for much lower number densities that for the snapshot case. These two cases are very similar in the luminosity range where they overlap, showing that the time averaging method does not change the predicted luminosity function significantly.

\section{Exploring the effect of varying parameters}
\label{app:varying_parameters}

\begin{figure*}
\centering
\includegraphics[width=.7\linewidth]{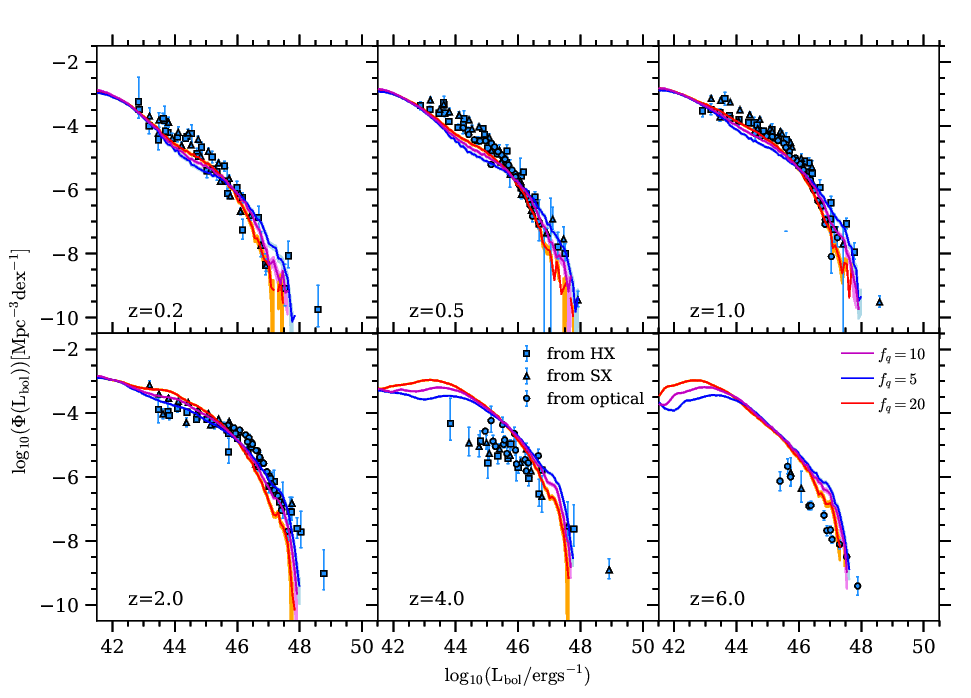}
\caption{Exploring the effect on the AGN bolometric luminosity function of varying the parameter $f_{\mathrm{q}}$. Shown are 
$f_{\mathrm{q}} = 5$ (blue), $f_{\mathrm{q}} = 10$ (purple, the fiducial model) and $f_{\mathrm{q}} = 20$ (red). The shading shows the Poisson errors 
of the distribution.}
\label{fig:bol_LF_varying_ftq} 
\end{figure*}

\begin{figure*}
\centering
\includegraphics[width=.7\linewidth]{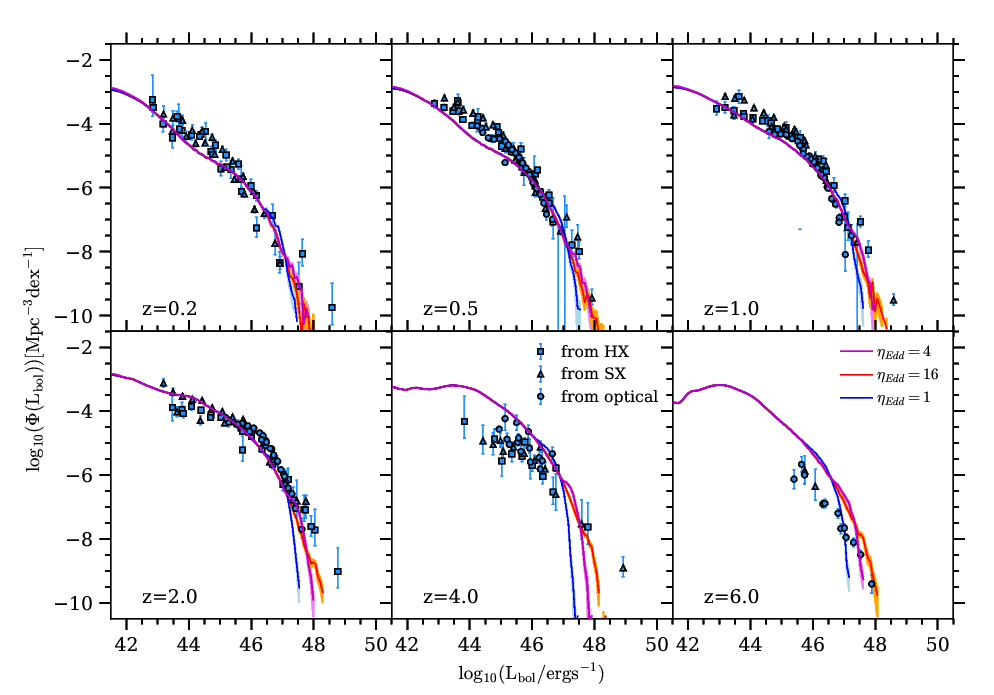}
\caption{Exploring the effect of varying $\eta_{\rm{Edd}}$. Shown are $\eta_{\rm{Edd}} = 1$ (blue), 
$\eta_{\rm{Edd}} = 4$ (purple, the fiducial model) and $\eta_{\rm{Edd}} = 16$ (red).}
\label{fig:bol_LF_varying_nEdd} 
\end{figure*}

\begin{figure*}
\centering
\includegraphics[width=.7\linewidth]{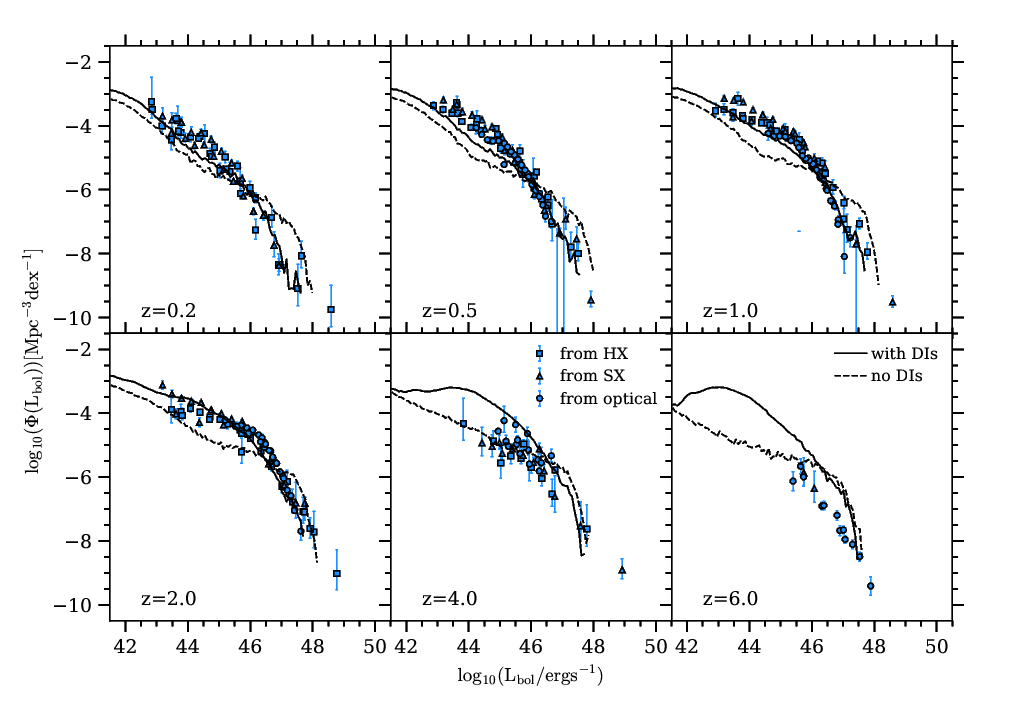}
\caption{Exploring the effect of switching off disc instabilities. Shown are the fiducial model (solid) 
and the model with disc instabilities switched off (dashed).}
\label{fig:L800_bol_stabledisk} 
\end{figure*}

\begin{figure*}
\centering
\includegraphics[width=.7\linewidth]{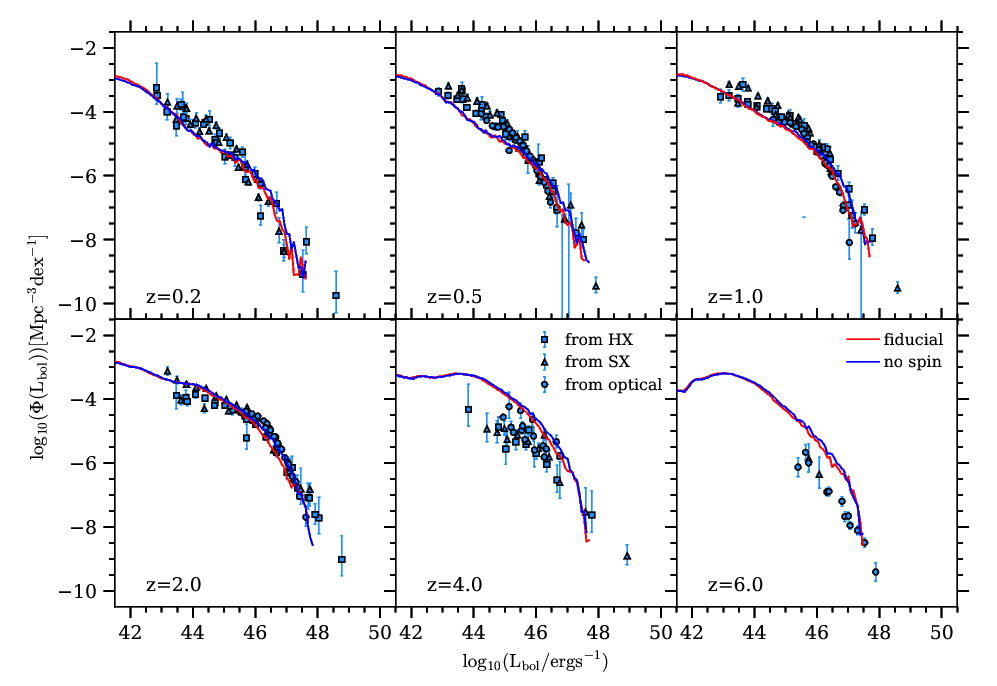}
\caption{Exploring the effect of turning off the SMBH spinup evolution: the model with chaotic mode accretion spinup and merger spinup
(red) and the model with no accretion nor merger spinup with a thin disc accretion efficiency, $\epsilon_{TD} = 0.1$ (blue).}
\label{fig:L800_bol_nospin} 
\end{figure*}

\begin{figure*}
\centering
\includegraphics[width=.7\linewidth]{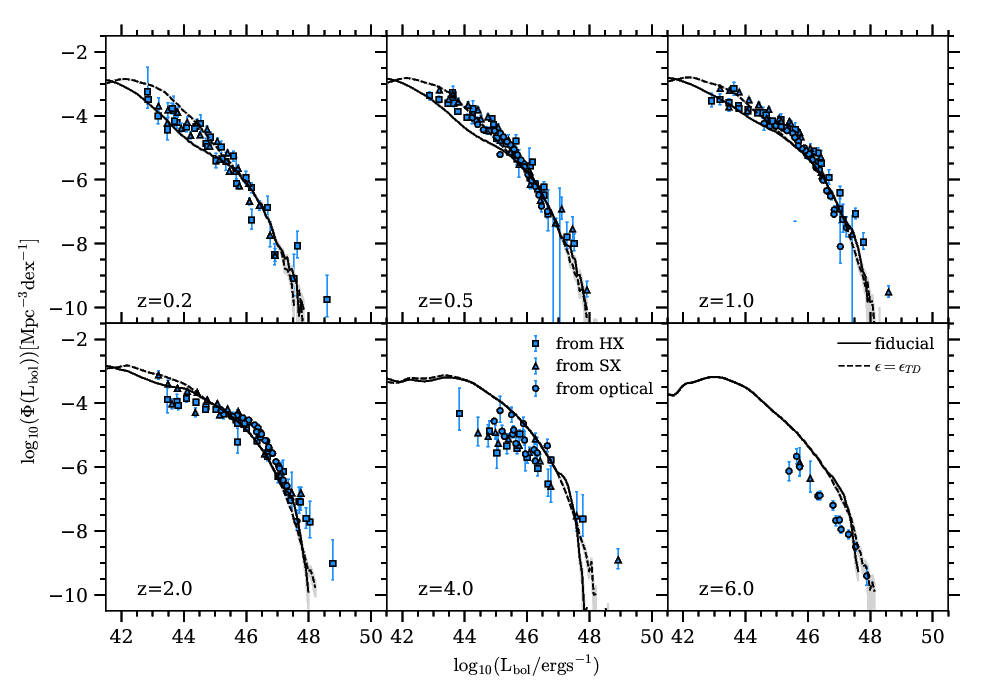}
\caption{Exploring the effect of changing the accretion efficiency $\epsilon$: the model with $\epsilon = \epsilon_{TD}$ as the 
accretion efficiency for all $\dot{m}$ regimes (black dashed) and the fiducial model (black solid).}
\label{fig:L800_bol_etd} 
\end{figure*}

We show the effect on the bolometric luminosity function of varying some of the free parameters for SMBH and AGN used in the model; in Figure \ref{fig:bol_LF_varying_ftq},
we show the effect of varying the parameter $f_{\mathrm{q}}$ (c.f. equation (\ref{eq:mdot_macc})). $f_{\mathrm{q}}$ affects the value of $\dot{M}$ and therefore the AGN 
luminosities. One expects a higher value of $f_{\mathrm{q}}$ to lead to lower values of $\dot{M}$ and therefore a steeper luminosity function
at the bright end, as we see in Figure \ref{fig:bol_LF_varying_ftq}. At the faint end, a lower value of $f_{\mathrm{q}}$ results in a poorer fit to the observations at low redshift ($z = 0.2,0.5,1$)
but is a better fit to the observations at high redshift ($z = 2,4,6$). At the bright end, a higher value of $f_{\mathrm{q}}$
seems to give a better fit to the observations at low redshift but gives a worse fit to the observations at 
high redshift (e.g. around $L_{\mathrm{bol}} \sim 10^{48} \mathrm{ergs^{-1}}$ at $z=4$). With these considerations in mind, we decide to keep the \cite{fani12}
value of $f_{\mathrm{q}} = 10$ for our predictions in this paper. 

We show the effect of varying the parameter $\eta_{\mathrm{Edd}}$ (c.f. equation (\ref{eq:lbol_se})) in Figure \ref{fig:bol_LF_varying_nEdd}.
$\rm{\eta_{Edd}}$ controls the suppression of the luminosity for super-Eddington accretion rates, where a low value of $\rm{\eta_{Edd}}$ corresponds to stronger 
luminosity suppression than a high value of $\eta_{\mathrm{Edd}}$. This parameter only affects the very bright
end of the luminosity function, as we would expect. This parameter also has more of an effect at high redshift,
where there are more super-Eddington sources. A value of $\eta_{\mathrm{Edd}} = 1$ gives a slightly better fit to the
bright end observations at $z=6$ but $\eta_{\mathrm{Edd}} = 16$ gives a better fit to bright end observations at 
$z=2$ and $z=4$. Therefore we once again to opt to keep the \cite{fani12}
value of $\rm{\eta_{Edd}} = 4$ for our predictions in this paper.

We show the effect of switching off disc instabilities in Figure \ref{fig:L800_bol_stabledisk}. We show the fiducial model alongside 
a model in which all discs are stable and so no disc
instability starbursts occur. Disc instabilities dominate the AGN luminosity function at $z>2$, and so this is the regime where we 
expect turning off disc instabilities to have the most effect. For $L_{\mathrm{bol}} < 10^{46} \mathrm{ergs^{-1}}$, at $z>2$ switching off disc instabilities
results in fewer starbursts and so there are fewer objects at these luminosities. For $L_{\mathrm{bol}} > 10^{46} \mathrm{ergs^{-1}}$, at $z>2$ 
the two models are similar - this is because if we switch off disc instabilities, galaxy mergers trigger the starbursts that would
have otherwise happened due to disc instabilities. At $z<2$, switching off disc instabilities makes the luminosity function less steep.

We show the effect of switching off the accretion and merger spinup in Figure \ref{fig:L800_bol_nospin}. The radiative accretion 
efficiency given to the black holes is $\epsilon=0.1$.
The luminosity functions for the two models are generally similar, although the fiducial model has a slightly lower
number density at high luminosities.

We show the effect of changing the assumptions for accretion efficiency, $\epsilon$, in Figure \ref{fig:L800_bol_etd}. We compare the fiducial
model to a model in which the accretion efficiency is the thin disc accretion efficiency for all values of the specific mass accretion 
rate, $\dot{m}$. Interestingly, this result provides a slightly better fit to the bolometric luminosity function,
particularly for $z<0.5$ and $L_{\mathrm{bol}}<10^{45}\mathrm{ergs^{-1}}$, where the fiducial model underpredicts the number 
density. This is the regime where ADAFs dominate the luminosity function, and so this test suggests that
a better fit to the observed AGN luminosity function might be obtained if 
the radiative accretion efficiency for ADAFs is higher than the values assumed in our standard model.


\bsp	
\label{lastpage}
\end{document}